\documentclass[sigconf,authorversion,nonacm]{acmart}

\usepackage{tikz}
\usetikzlibrary{patterns,positioning,fit,arrows.meta,backgrounds}
\tikzset{
    module/.style={%
        draw, rounded corners,
        minimum width=#1,
        minimum height=7mm,
        font=\sffamily
        },
    module/.default=2cm,
    >=LaTeX
}

\usepackage{multirow}

\usepackage{pgfplots}
\pgfplotsset{compat=1.9}
\usepgfplotslibrary{groupplots}

\usepackage{caption}
\usepackage{subcaption}
\usepackage{cleveref}
\usepackage{pifont}

\usepackage{listings}
\usepackage{algorithm}
\usepackage{algpseudocode}

\AtBeginDocument{%
  }

\setcopyright{acmcopyright}
\copyrightyear{2018}
\acmYear{2018}
\acmDOI{XXXXXXX.XXXXXXX}

\acmConference[Conference acronym 'XX]{Make sure to enter the correct
  conference title from your rights confirmation emai}{June 03--05,
  2018}{Woodstock, NY}
\acmPrice{15.00}
\acmISBN{978-1-4503-XXXX-X/18/06}

\newcommand{\squishlist}{
	\begin{list}{$\bullet$}
		{ \setlength{\itemsep}{1pt}
			\setlength{\parsep}{1pt}
			\setlength{\topsep}{2.5pt}
			\setlength{\partopsep}{0.5pt}
			\setlength{\leftmargin}{1em}
			\setlength{\labelwidth}{1em}
			\setlength{\labelsep}{0.6em}
		}
	}
	\newcommand{\squishend}{
	\end{list}
}

\newcommand{\stitle}[1]{\vspace*{0.4em}\noindent{\bf #1.\/}}

\newcommand{\name}{SPT}
\begin{document}

\title{SPT: Fine-Tuning Transformer-based Language Models Efficiently with Sparsification}


% 1
\author{Yuntao Gui}
\affiliation{
  \institution{The Chinese University of Hong Kong}
  \country{Hong Kong SAR, China}
}
\email{ytgui@cse.cuhk.edu.hk}

% 2
\author{Xiao Yan}
\affiliation{
  \institution{Southern University of Science and Technology}
  \country{China}
}
\email{yanx@sustech.edu.cn}

% 3
\author{Peiqi Yin}
\affiliation{
  \institution{The Chinese University of Hong Kong}
  \country{Hong Kong SAR, China}
}
\email{pqyin22@cse.cuhk.edu.hk}

% 4
\author{Han Yang}
\affiliation{
  \institution{The Chinese University of Hong Kong}
  \country{Hong Kong SAR, China}
}
\email{hyang@cse.cuhk.edu.hk}

% 5
\author{James Cheng}
\affiliation{
  \institution{The Chinese University of Hong Kong}
  \country{Hong Kong SAR, China}
}
\email{jcheng@cse.cuhk.edu.hk}

\renewcommand{\shortauthors}{Yuntao Gui et al.}

\begin{abstract}
  Transformer-based large language models (e.g., BERT and GPT) achieve great success, and \textit{fine-tuning}, which tunes a pre-trained model on a task-specific dataset, is the standard practice to utilize these models for downstream tasks. However, Transformer fine-tuning has \textit{long running time} and \textit{high memory consumption} due to the large size of the models. We propose the \name{} system to fine-tune Transformer-based models efficiently by introducing \textit{sparsity}. We observe that the memory consumption of Transformer mainly comes from storing attention weights for multi-head attention (MHA), and the majority of running time is spent on feed-forward network (FFN).
  Thus, we design the \textit{sparse MHA} module, which computes and stores only large attention weights to reduce memory consumption, and the \textit{routed FFN} module, which dynamically activates a subset of model parameters for each token to reduce computation cost.
  We implement \name{} on PyTorch and customize CUDA kernels to run sparse MHA and routed FFN efficiently. Specifically, we use product quantization to identify the large attention weights and compute attention via sparse matrix multiplication for sparse MHA.
  For routed FFN, we batch the tokens according to their activated model parameters for efficient computation. We conduct extensive experiments to evaluate \name{} on various model configurations. The results show that \name{} consistently outperforms well-optimized baselines, reducing the peak memory consumption by up to 50\% and accelerating fine-tuning by up to $2.2\times$.
\end{abstract}


\maketitle

% Introduction
\section{Introduction}
\label{sec:intro}

\begin{table}[!t]
	\small
	\centering
	\setlength{\tabcolsep}{1.75mm}
	\caption{Time and memory decomposition for a Transformer block  (see \S~\ref{sec:eval} for settings). \textit{Full} is full-parameter tuning, and LoRA is low-rank adaption. Total peak memory is smaller than summation due to dynamic tensor destruction.}
	\begin{tabular}{c|ccc|ccc}
		\toprule
		     & \multicolumn{3}{c|}{\textbf{Running Time}} & \multicolumn{3}{c}{\textbf{Peak Memory}}                                       \\
		\midrule
		     & MHA                                        & FFN                                      & Total    & MHA    & FFN    & Total  \\
		\midrule
		Full & 59.6 ms                                    & 128.8 ms                                 & 188.4 ms & 3.2 GB & 1.3 GB & 3.2 GB \\
		LoRA & 52.5 ms                                    & 108.5 ms                                 & 161.0 ms & 2.6 GB & 1.1 GB & 2.7 GB \\
		SPT  & 54.1 ms                                    & 54.9 ms                                  & 106.0 ms & 0.9 GB & 1.1 GB & 1.6 GB \\
		\bottomrule
	\end{tabular}
	\label{tab:motivation}
\end{table}

Recently, large language models (e.g., BERT~\cite{devlin2018bert}, RoBERTa~\cite{liu2019roberta}, GPT-2~\cite{radford2019gpt2}, OPT~\cite{zhang2022opt}, and LLaMa~\cite{touvron2023llama}) show amazing performance for many tasks such as machine translation, text generation, and question answering.
However, training these models from scratch is prohibitively expensive, requiring enormous general-purpose corpus and millions of GPU hours~\cite{narayanan21megatron, google23palm}, which are beyond reach for most practitioners. As such, \textit{fine-tuning} has become the standard practice to utilize these models for downstream tasks~\cite{houlsby19transfer, zaken22bitfit, hu2022lora, taori23alpaca}.
In particular, fine-tuning takes a pre-trained model, whose parameters encompass rich knowledge learned from a general-purpose corpus, and continues to train the model on a task-specific dataset, which is usually much smaller. It has been shown that fine-tuning yields good model quality for various tasks~\cite{chen2022adapter, zhang2023controlnet, taori2023alpaca, touvron2023llama2} while significantly reducing the cost of end-to-end training~\cite{houlsby19transfer, chen2022adaptformer, taori23alpaca}.

However, fine-tuning is still reasonably expensive given the large size of the language models, and many methods are proposed to improve efficiency such as BitFit~\cite{zaken22bitfit}, Input-Tuning~\cite{an22inputadapter}, and LoRA~\cite{hu2022lora}. Among these methods, \textit{low-rank adaptation}~(LoRA) is popular, which freezes the pre-trained model parameters and introduces a small number of new parameters to update during fine-tuning.
Specifically, LoRA uses small low-rank matrices as the new model parameters to reduce the number of parameters to update and is widely used due to its good model quality and efficiency~\cite{liu2022tuning, ding2022delta}. For instance,  Table~\ref{tab:motivation} shows that LoRA reduces both running time and memory consumption compared with full-parameter tuning that updates all the original model parameters.
Note that besides running time, memory consumption is also important for fine-tuning as the memory requirement scales with sequence length for language models, reduced memory consumption allows the models to be trained on a wider variety of downstream tasks (e.g., chatbot and text summarization).

% and reduced memory consumption allows to train with longer sequence, which usually yields better model quality by providing more context.

Although LoRA already outperforms full-parameter tuning, we ask the following research problem---\textit{can the efficiency of LoRA be improved further?} To explore the opportunities for improvement, we profile the costs of a Transformer block in Table~\ref{tab:motivation}.
Transformer is the basic building unit of large language models, and most models can be viewed as stacking multiple Transformer blocks. In a Transformer block, there are two main modules, i.e., \textit{multi-head attention} (MHA) and \textit{feed-forward network} (FFN). MHA allows the tokens in a sequence to attend to each other to capture context, while FFN transforms the token embeddings by matrix mapping. The results in Table~\ref{tab:motivation} show that for both full-parameter tuning and LoRA, the main memory consumption is caused by MHA while the majority of running time is spent on FFN.

Motivated by the observations above, we build the \name{} system to conduct LoRA fine-tuning efficiently by exposing sparsity in the two key modules of Transformer. In particular, we find that MHA takes huge memory to store the attention weights between all token pairs, and propose the \textit{sparse MHA}, which keeps only the top-$L$ attention weights for each token. This does not harm model quality because the attention weights are generated by softmax, and thus the top-$L$ weights dominate the total weights and yield a small approximation error. The FFN is computationally expensive because its projection matrices are large with many parameters. We propose the \textit{routed FFN}, which organizes the parameters in blocks and dynamically determines the parameter blocks to activate for each token. Routed FFN reduces computation as each token now uses only some of the parameters.

We implement \name{} on PyTorch and customize CUDA implementations for sparse MHA and routed FFN because PyTorch does not support their underlying sparse operators. In particular, to identify the top-$L$ attention weights for each token in MHA, we utilize the product quantization technique to quantize the tokens and measure the similarity between two tokens using their common codewords. This is efficient as it avoids computing and sorting floating point distances.
By focusing on only the top-$L$ attention weights, we can model the attention calculation as sparse matrix multiplications (e.g., SDDMM and SpMM) for efficiency.
The routed FFN causes an efficiency challenge because different tokens activate different model parameters, and thus it is difficult to batch their computation. To tackle this problem, we design a novel blocked sparse matrix-vector multiplication (BSpMV) procedure, which batches the tokens that activate the same block in the model parameters for efficient computation.

We conduct extensive experiments to evaluate \name{} and to compare with well-optimized full-parameter tuning and LoRA implementations.
For end-to-end fine-tuning on two billion-parameter models, \name{} achieves a maximum $1.47\times$ speedup with only marginal degradation to model quality.
Considering five popular Transformer blocks with different configurations (e.g., embedding dimension and projection matrix size), \name{} consistently outperforms full-parameter tuning and LoRA with $2.2\times$ speedup and uses 50\% less memory, in the best cases.
We also conduct micro experiments to validate our designs. The results show that the sparse MHA effectively reduces memory consumption, and the routed FFN effectively reduces running time. This can also be observed from Table~\ref{tab:motivation}, where \name{} yields a larger performance gain over LoRA than LoRA over full-parameter tuning.

To summarize, we make the following contributions.

\squishlist
\item We observe two key opportunities to improve LoRA fine-tuning, i.e., the memory consumption of MHA and running time of FFN.
\item By inducing sparsity, we propose the sparse MHA and routed FFN as two key modules to improve the efficiency of Transformer.
\item To run our sparse modules efficiently, we adopt a suite of techniques such as product quantization, bucket-sort-based ranking, and block-based computation batching.
\item We implement all these ideas in the \name{} system on PyTorch and make it open-source\footnote{GitHub repository: https://github.com/ytgui/SPT-proto} to benefit language model practitioners.
\squishend

% Background and Motivation
\section{Background}\label{sec:background}

\begin{figure}[!t]
	\centering
	\includegraphics[width=0.8\columnwidth]{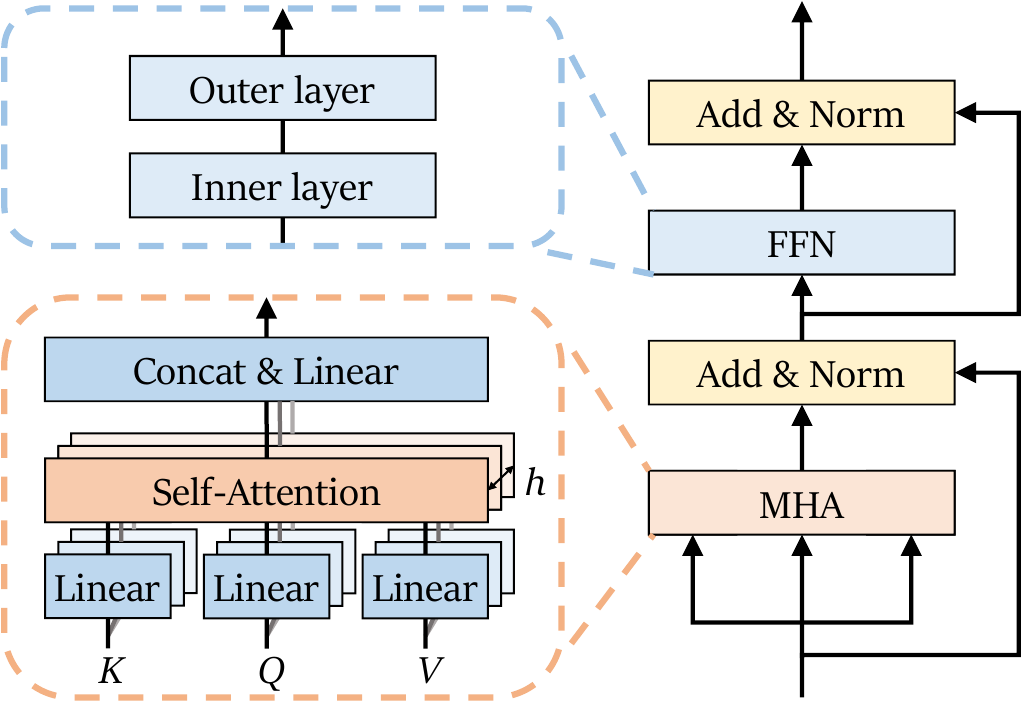}
	\caption{The structure of a Transformer block.}
	\label{fig:transformer}
\end{figure}

In this part, we introduce the basics of Transformer-based models and their fine-tuning to facilitate the subsequent discussions.

\subsection{Transformer Basics}\label{subsec:bg-transformers}

Transformer-based models (e.g., BERT~\cite{devlin2018bert}, GPT~\cite{radford2019gpt2}, OPT~\cite{zhang2022opt}, and LLaMA~\cite{touvron2023llama}) take a sequence of tokens (e.g., keywords in a paragraph) and produce an embedding vector for each token.
This is achieved by passing the initial token embeddings through several Transformer blocks. As shown in Figure~\ref{fig:transformer}, each Transformer block has two main components, i.e., \textit{multi-head attention} (MHA) and \textit{feed-forward network} (FFN).

MHA involves multiple attention heads that work in parallel, and an attention head allows each token to aggregate all tokens in the sequence based on correlation (called `attend to' the tokens).
In particular, the input of an attention head is $X \in \mathbb{R}^{n \times d}$, where $n$ is the sequence length and $d$ is the dimension of the embedding for each token divided by the total number of attention heads (called embedding dimension afterward).
Each row $x_i \in \mathbb{R}^{d}$ of $X$ corresponds to the input of a token to one attention head. The attention head uses learnable projection matrices $W^Q$, $W^K$, and $W^V$ to project the input $X$ to obtain query matrix $Q$, key matrix $K$, and value matrix $V$ as follows
\begin{equation}
	Q=X W^Q, \quad K=X W^K,\quad V  =  X W^V.
\end{equation}
Then, the $Q$, $K$, and $V$ matrices are used to compute output as
\begin{equation}\label{equ:attention}
	Y=\mathsf{Attention}(Q, K, V)  =  \mathsf{Softmax}(QK^T)V.
\end{equation}
In particular, Eq.~\eqref{equ:attention} uses $Q$ and $K$ to compute an attention weight matrix $A=\mathsf{Softmax}(QK^T)\in \mathbb{R}^{n \times n}$, where an entry $a_{ij}\in A$ models the correlation between token $i$ for token $j$. The final output is $Y=AV$, i.e., each row $y_i \in \mathbb{R}^{d}$ of the output matrix $Y \in \mathbb{R}^{n \times d}$ is a weighted sum of the rows in $V$. More specifically, $y_i$ can be expressed as
\begin{equation}\label{equ:MHA single}
	y_i=\sum_{j=1}^{n}\frac{\exp(q_i^{T} k_j)}{\sum_{j=1}^{n}\exp(q_i^{T} k_j)}v_j,
\end{equation}
where $a_{ij}=\exp(q_i^{T} k_j)/(\sum_{j=1}^{n}\exp(q_i^{T} k_j))$ is the attention weight of token $i$ for token $j$ produced by the softmax function and we can simplify Eq.~\eqref{equ:MHA single} as  $y_i=\sum_{j=1}^{n}a_{ij}v_j$.

The FFN stacks two linear mapping layers to map the embeddings, which are called inner projection and outer projection, respectively. That is,
\begin{equation}\label{equ:FFN}
	H=\mathsf{ReLU}(XW_I), \quad Y=HW_O,
\end{equation}
where $W_I\in \mathbb{R}^{d \times D}$ and $W_O \in \mathbb{R}^{D \times d}$ are the parameters of the inner and outer projection matrices. The row dimension of $W_I$ (i.e., $d$) matches the column dimension of $X$, and $D$ is the intermediate dimension of FFN (usually larger than $d$). Note that MHA and FFN have different inputs and outputs although we use the same notations (i.e., $X$ and $Y$) for conciseness.

\subsection{Fine-tuning Transformer}\label{subsec:bg-tuning}

Due to the high computation cost and massive size of general-purpose corpus, training Transformer-based models from scratch is prohibitively expensive.
As a result, the standard practice when applying these models to a specific task is to fine-tune a pre-trained model.
In particular, fine-tuning takes a pre-trained model and continues to train the model on a dataset collected for the target task (e.g., chatbot for makeup sales).
Fine-tuning benefits from the substantial knowledge encoded in the parameters of the pre-trained model, while only required to train on a small task-specific dataset rather than a large general-purpose corpus.

During fine-tuning, one may directly update all the model parameters, which is called \textit{full-tuning}. However, full-tuning is expensive because there are many model parameters to update. As such, adapter-based fine-tuning is proposed, which freezes the pre-trained model parameters and introduces a small number of new parameters that are updated to adapt to the target task~\cite{chen2022adaptformer}.
The most popular adapter-based fine-tuning method is \textit{low-rank adaptation} (LoRA)~\cite{hu2022lora}. Consider a linear projection of the form $ Y = XW$ (which serves as the basic unit for both MHA and FFN), where $X \in \mathbb{R}^{n \times d}$ is the input and $W \in \mathbb{R}^{d \times h}$ is the model parameter, LoRA describes the projection as
\begin{align}
	Y & = XW + XBC=X(W+BC),
\end{align}
where $W$ contains the pre-trained parameters and is kept fixed while $B$ and $C$ are the new parameters and are updated during training. Moreover, both $B \in \mathbb{R}^{d \times r}$ and $C \in \mathbb{R}^{r \times h}$ are low-rank matrices with $r << d$, and thus the number of parameters to update is small for efficient gradient computation and update. LoRA is shown to yield comparable model accuracy to full-tuning. Inference with LoRA model is as efficient as the original model because we can merge the pre-trained and new model parameters as $W'=W+BC$ after fine-tuning. Thus, we focus on improving LoRA fine-tuning in this paper.

% Overview
\section{\name{} System Overview}
\label{sec:overview}

\begin{figure}[!t]
\centering
\includegraphics[width=1.0\columnwidth]{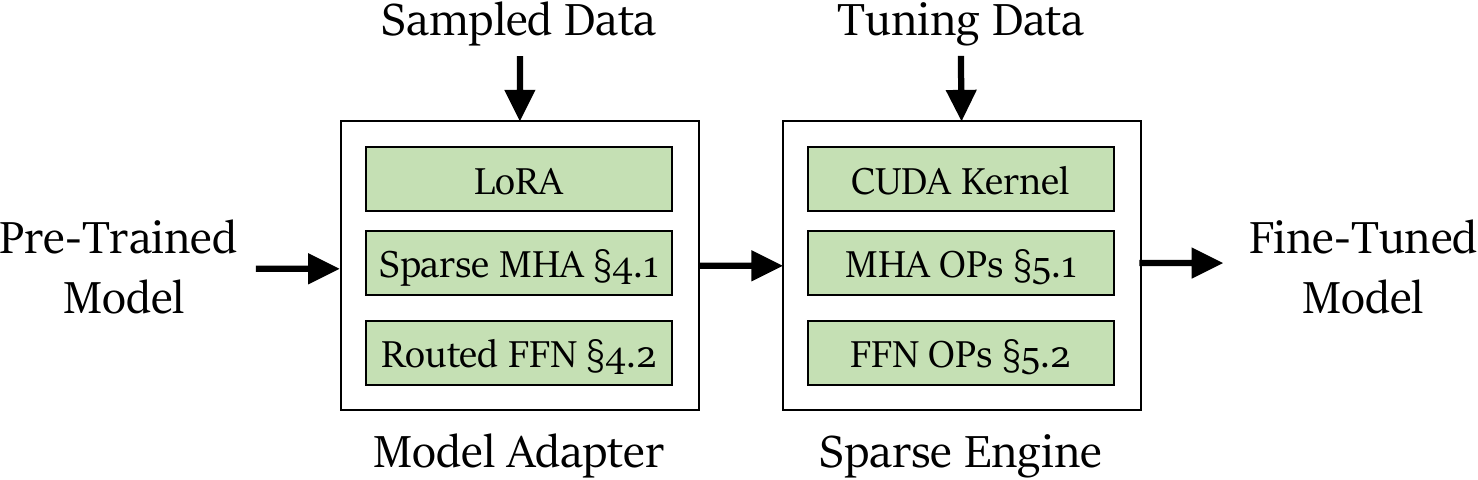}
\caption{The workflow of our \name{} system.}
\label{fig:sys}
\end{figure}

Our \name{} integrates algorithmic and systematic innovations to improve the efficiency of LoRA-based fine-tuning in terms of both running time and memory usage.
Note that reducing memory usage, in addition to running time, is also critical when tuning Transformer models because peak memory scales with sequence length. By lowering memory demands, the same GPU can handle longer sequences with a fixed memory capacity; longer sequences provide more context for language models and enable more downstream tasks~\cite{beltagy2020longformer,taori2023alpaca}.
As shown in \S\ref{sec:intro}, for LoRA fine-tuning, the main memory consumption comes from storing the attention matrix $A\in \mathbb{R}^{n \times n}$ for MHA while the main running time is consumed by FFN for the projections $Y=XW$. From the algorithm perspective, we observe that \textit{sparsification} effectively reduces both costs, i.e., memory consumption drops by keeping only some of the attention weights in $A$, and computation becomes faster by using only some of the parameters in $W$ to conduct projection.
Thus, we propose the sparse MHA and FFN modules to expose sparsity. From the system perspective, we build an efficient execution engine to run the sparse MHA and FFN modules on GPU, which is radically different from existing engines that use dense neural networks. Figure~\ref{fig:sys} provides an overview of \name{}, and we briefly introduce its workflow as follows.

\stitle{Model Adapter}
It prepares a pre-trained Transformer-based model for sparse fine-tuning. First, the pre-trained model parameters are loaded, and then the LoRA layers and related new parameters are inserted into the computation graph of the model.
It then replaces the standard MHA and FFN modules with our \textit{sparse MHA} (cf. \S\ref{subsec:sparse MHA}) and \textit{routed FFN} (cf. \S\ref{subsec:routed FFN}).

In particular, sparse MHA replaces the dense attention computation $A=QK^T$ in Eq.~\eqref{equ:attention} with sparse dense-dense matrix multiplication (SDDMM) $A'=SDDMM(Q, K^T)$, which computes only the top-$L$ largest attention weights for each row $q_i$ of the query matrix $Q$. As such, the sparse attention matrix $A'$ has only $n\times L$ non-zero entries and takes much smaller space than the $n\times n$ dense attention matrix $A$.
The output of sparse MHA is computed as $Y=A'V$ instead of $Y=AV$ in Eq.~\eqref{equ:attention}, and this is conducted by a sparse matrix-matrix multiplication (SpMM) $Y=SpMM(A',V)$.
The routed FFN induces sparsity by using only some rows of the inner projection matrix $W_I$ and columns of the outer projection matrix $W_O$.
This reduces the computation workloads (i.e., $XW_I$ and $HW_O$) in Eq.~\eqref{equ:FFN} for dense matrix multiplications.
Note that instead of permanently removing some parameters for all tokens (which harms model quality), our routed FFN dynamically determines the parameters to prune for each token.

\stitle{Sparse Engine}
\name{} implements an efficient computation engine to conduct fine-tuning on the modified computation graph.
The engine is built on PyTorch~\cite{fb17pytorch} and includes customized CUDA implementations for the sparse operators, which are not supported by PyTorch.
For sparse MHA, the key challenges are finding the top-$L$ keys for each query and conducting SDDMM and SpMM for sparse attention efficiently. We adapt the product quantization method to map the queries and keys and propose a bucket-sort-based procedure for top-$L$ selection; we also carefully design the index encoding for the sparse matrices to make SDDMM and SpMM efficient (cf. \S\ref{subsec:sys-mha}). For routed FFN, the key challenge is that different tokens activate different model parameters, which invalidates efficient batched computation. We design a novel blocked sparse matrix-vector multiplication (BSpMV) procedure, which batches the tokens that activate the same block in the model parameters for efficient computation (cf. \S\ref{subsec:sys-ffn}).

\name{} allows users to trade-off between training efficiency and model quality by configuring the strength of sparsification. This is done by setting $L$ (i.e., the number of attention weights to consider for each query vector in MHA) and $\beta$ (i.e., the portion of model parameters to use for FFN).
As we will show by experiments in \S\ref{sec:eval}, \name{} can significantly improve efficiency with only a marginal degradation to model quality. To help users determine the strength of sparsification, \name{} allows users to conduct short training trials on some sample data.

%\yt{TODO: describe sample data before tuning}
%
%\xiao{the engine part is too short, which shows a mismatch to the adapter part; can highlight the challenges of each part, and then give an outline of the solution}

% System Design
\section{The Sparse Transformer}\label{sec:algorithm}

\begin{figure}[!t]
  \centering
  \footnotesize
  \begin{subfigure}[t]{0.3\textwidth}
    \begin{tikzpicture}
      \begin{axis}[
          axis lines=left,
          xlabel=Propotion,
          ylabel=Cumulative weights,
          xmin=0.0,
          xmax=1.05,
          ymin=0.0,
          ymax=1.10,
          yticklabel style={rotate=90},
          scale only axis=true,
          width=0.8\textwidth,
          height=0.5\textwidth,
          legend cell align={center},
          legend style={anchor=south west, at={(0.4,0.1)}, legend columns=2}
        ]
        % w
        \addplot[mark=none, smooth] coordinates {
            (0.00,0.00) (0.0625,0.74) (0.125,0.88) (0.1875,0.92) (0.25, 0.95)
            (0.3125,0.96) (0.375,0.98) (0.4375,0.99) (0.5,1.0) (0.5625,1.0)
            (0.625,1.0) (0.6875,1.0) (0.75,1.0) (0.8125,1.0) (0.875,1.0)
            (0.9375,1.0) (1.0,1.0)
          };
        % baseline
        \addplot[dotted, style=thick] coordinates {
            (0.0,0.9) (1.0,0.9)
          };
%        \legend{\scriptsize{softmax}}
      \end{axis}
    \end{tikzpicture}
\end{subfigure}
  \caption{CDF of the softmax attention weights in MHA.}
  \label{fig:cdf}
\end{figure}
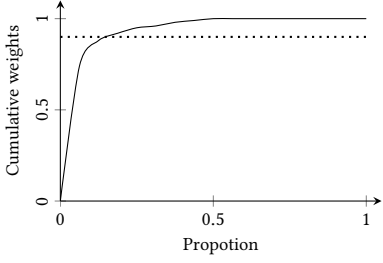

% fc1
% w_size: torch.Size([10240, 2560])
% sigma_size: torch.Size([2560])
% 0.50 sigma: 926
% 0.75 sigma: 1597
% 0.80 sigma: 1755
% 0.90 sigma: 2109
% 0.95 sigma: 2313

% fc2
% w_size: torch.Size([2560, 10240])
% sigma_size: torch.Size([2560])
% 0.50 sigma: 887
% 0.75 sigma: 1550
% 0.80 sigma: 1711
% 0.90 sigma: 2079
% 0.95 sigma: 2296

% x
% x_size: torch.Size([512, 2560])
% sigma_size: torch.Size([512])
% 0.50 sigma: 108
% 0.75 sigma: 237
% 0.80 sigma: 273
% 0.90 sigma: 363

% h
% x_size: torch.Size([512, 10240])
% sigma_size: torch.Size([512])
% 0.50 sigma: 74
% 0.75 sigma: 202
% 0.80 sigma: 240
% 0.90 sigma: 338

% y
% x_size: torch.Size([512, 2560])
% sigma_size: torch.Size([512])
% 0.50 sigma: 118
% 0.75 sigma: 244
% 0.80 sigma: 278
% 0.90 sigma: 364

%\xiao{For figure 3; we may use two models, such that softmax and FFN can have a separate figure, each with two subplots}

In this part, we introduce our two key sparse modules for Transformers, i.e., sparse multi-head attention (MHA) and routed feed-forward network (FFN).

\subsection{Sparse Multi-head Attention}\label{subsec:sparse MHA}

\stitle{Sparse attention}
We observe that the output of an attention head can be approximated by considering only the top-$L$ attention weights for each query vector $q_i$.
Recall that the $i\textsuperscript{th}$ output embedding (i.e., the $i\textsuperscript{th}$ row of the output matrix $Y$) of MHA can be expressed as
\begin{equation*}
y_i=\sum_{j=1}^{n}a_{ij}v_j,
\end{equation*}
where attention weight $a_{ij}=\exp(q_i^{T} k_j)/(\sum_{j=1}^{n}\exp(q_i^{T} k_j))$ is computed using the softmax function, which normalizes the scores $\exp(q_i^{T} k_j)$ across all keys $k_j$ for each query $q_i$. This gives a probability distribution over the keys for each query as softmax ensures that the attention weights for each query sum to 1.
The relative weight $a_{ij'}/a_{ij}=\exp(q_i^T k_{j'}-q_i^T k_j)$ depends only on the relative scores $q_i^T k_{j'}-q_i^T k_j$ for two keys $j'$ and $j$.
If $q_i^T k_{j'}$ is moderately larger than $q_i^T k_j$, $a_{ij'}/a_{ij}$ will be large as function $\exp(x)$ increases quickly with $x$.
Thus, for each query $q_i$, the keys with the highest dot product score $q_i^T k_j$ will dominate the attention weights.

Figure~\ref{fig:cdf} shows this phenomenon by plotting the cumulative distribution function (CDF) of the attention weights, where the top-15\% attention weights take up 90\% of the total attention weights.
As attention weights beside the top-$L$ are small, we can approximate $y_i$ as $y'_i=\sum_{j\in\mathcal{S}_k} a_{ij} v_j$, where ${S}_k$ denotes the set of keys with the top-$L$ scores for $q_i$.
We also revise softmax as $a_{ij}=\exp(q_i^{T} k_j)/(\sum_{j\in\mathcal{S}_k}\exp(q_i^{T} k_j))$ such that the attention weights of the top-$L$ keys sum to 1. This approximation makes the attention matrix sparse to save memory consumption as most attention weights are treated as 0.

\begin{figure}[!t]
\centering
\setlength{\abovecaptionskip}{0.25cm}
\includegraphics[width=0.99\columnwidth]{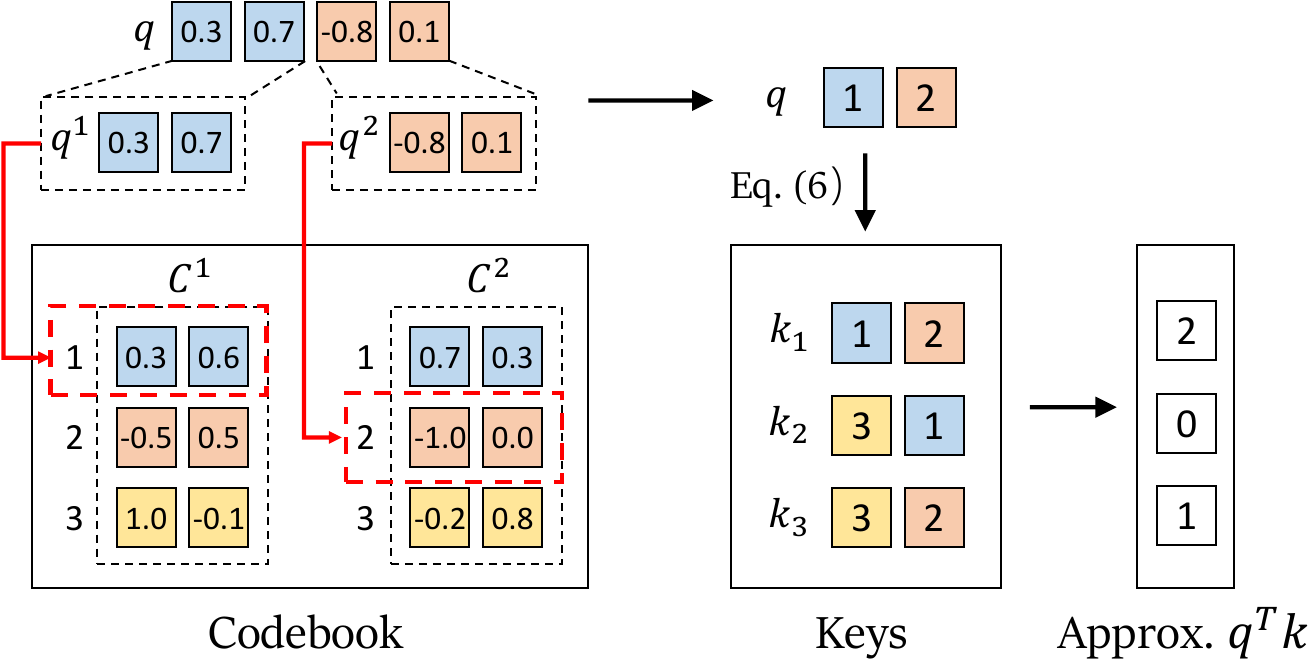}
\caption{An illustration of PQ.}
\label{fig:pq-sample}
\end{figure}

\stitle{PQ for efficient sparsification}
Identifying the top-$L$ attention scores for $q_i$ is essentially the maximum inner product search (MIPS) problem, which, given a query vector $q$, finds the top-$L$ vectors having the maximum inner product with $q$ in a dataset containing $n$ vectors~\cite{shen15mips, tiwari22mips}.
A naive solution will first compute the scores for all keys and then identify the top-$L$. This introduces high computation costs.
Pre-indexing $k_{j}$, as vector databases do for efficient top-$L$ retrieval, with techniques such as proximity graphs and trees, is also not feasible, because the keys are generated dynamically by MHA.
Fortunately, there are many approximate solutions to MIPS, which find the most (rather than all) of the top-$L$ to trade for efficiency~\cite{johnson19billion}.
We utilize the \textit{product quantization} (PQ) technique as our approach because it is computationally lightweight and can run efficiently on GPU.

PQ quantizes a vector $q \in \mathbb{R}^{d}$ using $M$ codebooks $\mathcal{C}^1,\mathcal{C}^2,\cdots,\mathcal{C}^M$.
Each codebook $\mathcal{C}^m$ contains $E$ codewords and each codeword is a $d'=d/M$-dimensional vector, i.e., $\mathcal{C}^m=\{c^m[1],c^m[2],\cdots,c^m[E]\}$ with $c^m[e]\in\mathbb{R}^{d'}$ for $1 \le m \le M$ and $1 \le e \le E$.
\footnote{We assume that $d$ is divisible by $M$. If this does not hold, PQ usually allocates fewer dimensions to the last codebook. This subtlety does not affect our discussions.}
Vector $q$ is first chopped into $M$ sub-vectors $[q^1,q^2,\cdots,q^M]$ with each sub-vector $q^m\in \mathbb{R}^{d'}$.
Then, each sub-vector $q^m$ of $q$ is approximated by finding the codeword with the minimum Euclidean distance to it in codebook $\mathcal{C}^m$.
Denote $t^m_q$ as the index of such codeword in codebook $\mathcal{C}^m$, $q$ is approximated by concatenating the codewords from each codebook $\tilde{q}=[c^1[t^1_q],c^2[t^2_q],...,c^M[t^M_q]]$.
Figure~\ref{fig:pq-sample} provides a running example of PQ with 2 codebooks and each codebook containing 3 codewords. In particular, $q$ finds codeword $t^1_q=1$ in codebook $\mathcal{C}^1$ and codeword $t^2_q=2$ in codebook $\mathcal{C}^2$, and thus $\tilde{q}=[c^1[1],c^2[2]]=[0.3,0.6,-1.0,0.0]$.

With PQ, the query-key inner product $q^{T}k$ can be approximated as $\tilde{q}^{T}\tilde{k}$ using the quantized representations of the query and key.
This approximation is typically computed by looking up an inner product table for each codebook and summing up the terms from all codebooks as $\tilde{q}^{T}\tilde{k}=\sum_{m=1}^{M} c^m[t^m_q]^{T} c^m[t^m_k]$, where $c^m[t^m_q]^{T} c^m[t^m_k]$ is pre-computed for all codeword combinations.
However, performing both this summation and the subsequent top-$L$ identification with floating point numbers is expensive. We propose conducting similarity computation with integers for improved efficiency.
Specifically, given that $\tilde{q}=[c^1[t^1_q],c^2[t^2_q],...,c^M[t^M_q]]$ and $\tilde{k}=[c^1[t^1_k],c^2[t^2_k],...,c^M[t^M_k]]$, we approximate $q^{T}k$ as
\begin{equation}\label{equ:count}
s(q,k)=\sum_{m=1}^{M} \mathbb{I}[t^m_q=t^m_k],
\end{equation}
where $\mathbb{I}[\cdot]$ is the indicator function. The intuition is that if $q$ and $k$ are mapped to the same codeword in more codebooks, they are more likely to have a large inner product. A running example is provided in the right part of Figure~\ref{fig:pq-sample}, where the codeword indexes of $q$ are compared to 3 keys.
As both computing and ranking the scores involve only integers, we adopt an efficient bucket-sort-based algorithm in \S\ref{sec:system} for top-$L$ selection. This approximation also provides good accuracy and the recall rate of identifying top-$L$ is close to 90\%, which is essential for good model quality.
%\yt{(To Xiao: I made a lot changes in this section.)}

\begin{algorithm}[t!]
\caption{The procedure of sparse MHA.}
\label{alg:sparse-mha}
\begin{flushleft}
\textbf{Input:} the query, key, and value embeddings $Q,\ K,\ V$ \\
\textbf{Output:} attention output $Y$
\end{flushleft}
\begin{algorithmic}[1]
\State Quantize the query vectors $C_Q \gets \mathsf{quantize}(Q)$
\State Quantize the key vectors $C_K \gets \mathsf{quantize}(K)$
\State Select the top-$L$ keys for each query $\tilde{Q},\ \tilde{K} = \mathsf{select}(C_Q,\ C_K)$
\State Compute sparse attention weight $\tilde{A} \gets \mathsf{softmax}( \tilde{Q}\tilde{K}^T )$
\State Compute attention output $Y \gets \tilde{A}V$
\end{algorithmic}
\end{algorithm}

\stitle{Workflow}
Algorithm~\ref{alg:sparse-mha} summarizes the procedure of our sparse MHA. First, the query and key vectors are quantized by the PQ codebooks (line 1 and 2). Second, we identify the top-$L$ query-key pairs according to the PQ codewords using Eq.~\eqref{equ:count} (line 3). Then, we compute the sparse attention matrix according to the top-$L$ query-key pairs (line 4) and the final output (line 5).
Here line 4 and line 5 involve sparse matrices instead of dense matrices as in the original MHA, and thus we develop specialized implementations in \S\ref{sec:system}.
Note that Transformers can be categorized into Encoders or Decoders, distinguished by whether there is a look ahead mask to prevent attending to future tokens. \name{} can be utilized in both the Encoder and Decoder architectures, we achieve this by applying the look ahead mask when computing softmax.

\stitle{Complexity analysis}
Assuming that the sequence length is $n$ and the embedding dimension is $d$ for a single head, the original attention has time complexity $O(n^2 d)$ to compute all scores and space complexity $O(n^2)$ to store the attention weights. For PQ with $M$ codebooks each containing $E$ codewords, the time to quantize the keys and queries is $O(ndE)$; the time to select the top-$L$ keys for all queries is $O(n^2)$; and finally computing the top-$L$ attentions take time $O(nLd)$; thus the overall time complexity is $O(ndE+n^2+nLd)$.
The space complexity is $O(nL)$ to store the attention weights and $O(Ed)$ to store the codebooks. We set the number of codewords $E$ as a small number (e.g., 16 and 32, which allows ignoring the terms involving $E$), the value of $L$ in top-$L$ as $\lambda n$. Thus, the space complexity of sparse MHA is roughly $O(\lambda n^2)$.
For instance, $\lambda=0.25$ yields a memory saving of 75\%.

%Note that although sparse MHA reduces the theoretical time complexity, the actual running time does not decrease because the original dense matrix operations are more efficient than our sparse matrix operations. As such, we mainly use sparse MHA to reduce memory consumption.
%\yt{(TODO: to show the ideal space complexity cannot be achieved due to CSR encoding)}

\subsection{Routed Feed-froward Network}\label{subsec:routed FFN}

To reduce the computation cost of FFN, we prune a portion of its parameters, which means that only a subset of the parameters are activated for computation.
Weight pruning techniques are widely used to improve efficiency for neural network models~\cite{wang20pruning}, and they fall into two main categories, i.e., static pruning and dynamic pruning.
Static pruning removes weights from the model based on their importance (e.g., magnitude) and does not consider the input. Dynamic pruning makes the pruning input-aware by removing different weights for different inputs.

\begin{figure}[!t]
  \centering
  \footnotesize
  \begin{subfigure}[t]{0.3\textwidth}
    \begin{tikzpicture}
      \begin{axis}[
          axis lines=left,
          xlabel=Propotion,
          ylabel=Cumulative singular value,
          xmin=0.0,
          xmax=1.05,
          ymin=0.0,
          ymax=1.10,
          yticklabel style={rotate=90},
          scale only axis=true,
          width=0.8\textwidth,
          height=0.5\textwidth,
          legend cell align={center},
          legend style={anchor=south west, at={(0.55,0.1)}, legend columns=1}
        ]
        % w
        \addplot[mark=none, smooth] coordinates {
            (0.00,0.00) (0.0625,0.12) (0.125,0.21) (0.1875,0.30) (0.25, 0.37)
            (0.3125,0.45) (0.375,0.51) (0.4375,0.58) (0.5,0.64) (0.5625,0.70)
            (0.625,0.76) (0.6875,0.80) (0.75,0.85) (0.8125,0.89) (0.875,0.93)
            (0.9375,0.97) (1.0,1.0)
          };
        % x
        \addplot[mark=*, mark size=1pt, smooth] coordinates {
            (0.00,0.04) (0.0625,0.26) (0.125,0.38) (0.1875,0.47) (0.25,0.55)
            (0.3125,0.62) (0.375,0.68) (0.4375,0.73) (0.5,0.78) (0.5625,0.82)
            (0.625,0.86) (0.6875,0.89) (0.75,0.92) (0.8125,0.95) (0.875,0.97)
            (0.9375,0.99) (1.0, 1.0)
          };
        % h
        \addplot[mark=o, mark size=1pt, smooth] coordinates {
            (0.0,0.13) (0.0625,0.36) (0.125,0.47) (0.1875,0.56) (0.25,0.63)
            (0.3125,0.69) (0.375,0.74) (0.4375,0.78) (0.5,0.82) (0.5625,0.85)
            (0.625,0.88) (0.6875,0.91) (0.75,0.94) (0.8125,0.96) (0.875,0.97)
            (0.9375,0.99) (1.0,1.0)
          };
        % baseline
        \addplot[dotted, style=thick] coordinates {
            (0.0,0.5) (1.0,0.5)
          };
        \legend{\scriptsize{$W_I$}, \scriptsize{$X$}, \scriptsize{$H$}}
      \end{axis}
    \end{tikzpicture}
    \end{subfigure}
  \caption{CDF of the normalized singular values in FFN.}
  \label{fig:cdf-ffn}
\end{figure}
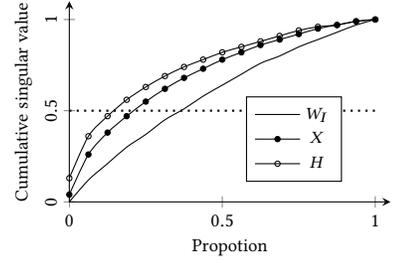

% fc1
% w_size: torch.Size([10240, 2560])
% sigma_size: torch.Size([2560])
% 0.50 sigma: 926
% 0.75 sigma: 1597
% 0.80 sigma: 1755
% 0.90 sigma: 2109
% 0.95 sigma: 2313

% fc2
% w_size: torch.Size([2560, 10240])
% sigma_size: torch.Size([2560])
% 0.50 sigma: 887
% 0.75 sigma: 1550
% 0.80 sigma: 1711
% 0.90 sigma: 2079
% 0.95 sigma: 2296

% x
% x_size: torch.Size([512, 2560])
% sigma_size: torch.Size([512])
% 0.50 sigma: 108
% 0.75 sigma: 237
% 0.80 sigma: 273
% 0.90 sigma: 363

% h
% x_size: torch.Size([512, 10240])
% sigma_size: torch.Size([512])
% 0.50 sigma: 74
% 0.75 sigma: 202
% 0.80 sigma: 240
% 0.90 sigma: 338

% y
% x_size: torch.Size([512, 2560])
% sigma_size: torch.Size([512])
% 0.50 sigma: 118
% 0.75 sigma: 244
% 0.80 sigma: 278
% 0.90 sigma: 364

\begin{figure}[!t]
\centering
\setlength{\abovecaptionskip}{0.25cm}
\includegraphics[width=0.80\columnwidth]{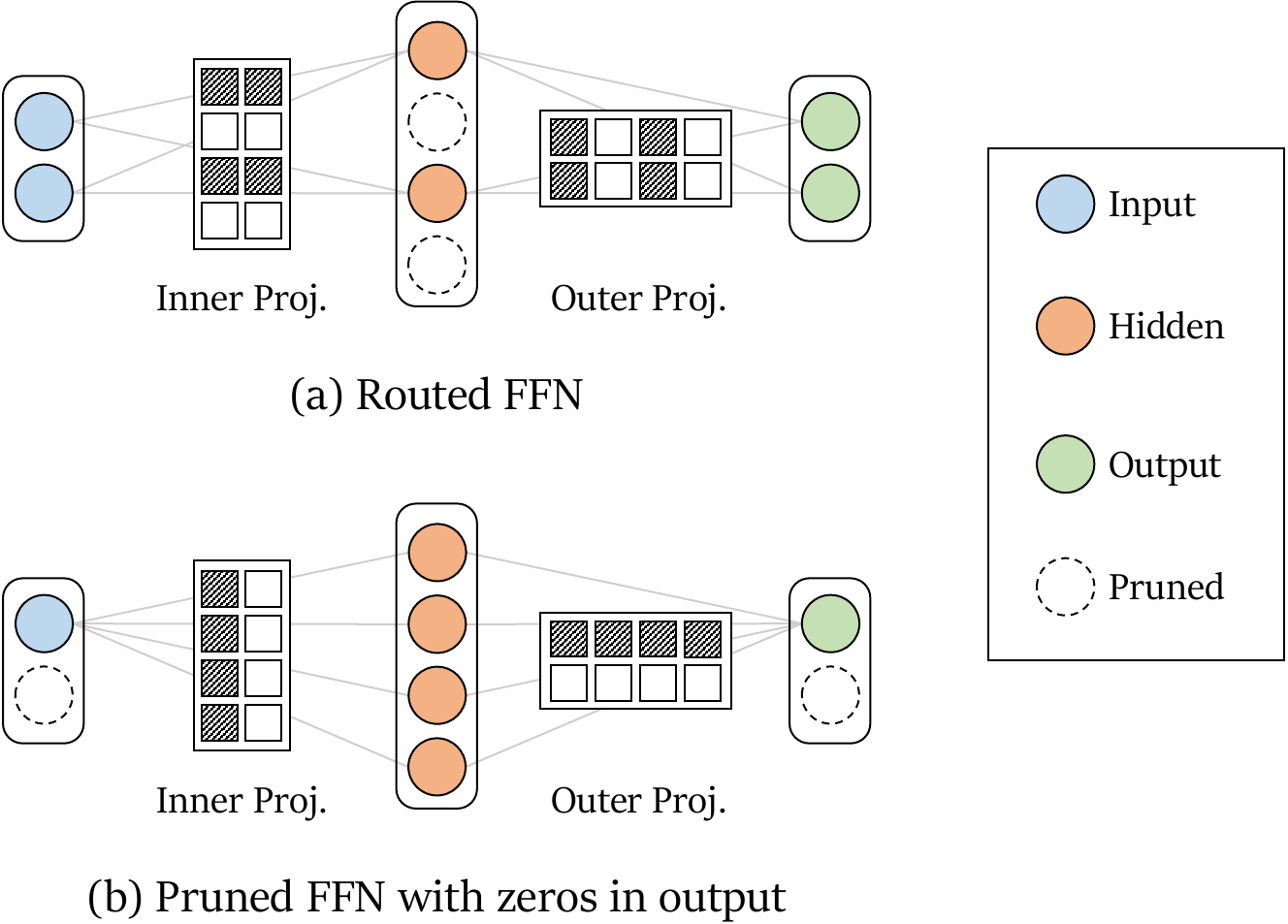}
\caption{An illustration of the routed FFN.}
\label{fig:routed-ffn}
\end{figure}

We conducted an empirical study to determine whether FFN should use static pruning or dynamic pruning.
Figure~\ref{fig:cdf-ffn} plots the CDF of the singular values of the inner projection matrix, feature matrices before and after the projection (the \textit{ReLU} activation is not considered).
Specifically, we use the FFN in the last Transformer block of the OPT-2.7B model and set the sequence length as 512 without padding.
Figure~\ref{fig:cdf-ffn} shows that the inner projection matrix (i.e., $W_I$) has a high-rank, and the cumulative singular value increases almost linearly with the number of singular values.
This indicates that the projection matrix utilizes most of its model capacity (i.e., small redundancy).
However, the projection output (i.e., $H$) has a relatively low-rank, with the top 25\% singular values taking up more than 50\% of the energy. Similar observations are also made on other FFN blocks and different Transformer architectures.
These results suggest that static pruning on weight matrices will harm model capacity and degrade quality, while the projection output has a low-rank and can be pruned by inducing sparsity.
As such, we select the dynamic pruning strategy, which keeps the capacity of $W_I$ and derives sparsity on $H$.
%\yt{(fix notations in this paragraph and the CDF plot)}

\stitle{Dynamic routing}
The question becomes how to conduct dynamic pruning while still supporting efficient computation. The naive solution is to prune the individual entries in the projection matrices. However, this induces highly irregular sparsity (as in our sparse MHA) and makes it inefficient to run the sparsified model.
As such, we use a block-based pruning strategy, which organizes adjacent rows and columns of the projection matrices as a block and conducts pruning at block granularity. This means that entries inside a block are activated together and can be handled by dense matrix computation effectively on GPU.

In particular, for an FFN whose inner projection is $h=\mathsf{ReLU}(xW_I)$ and outer projection is $y=\mathsf{ReLU}(xW_O)$, we prune the rows of the inner projection matrix and the columns of the outer projection matrix as shown in Figure~\ref{fig:routed-ffn}a.
As can be observed from Figure~\ref{fig:routed-ffn}a, pruning the row of $W_I$ sets some entries of the indeterminate embedding $h$ to zero, which means that the corresponding columns of $W_O$ will not be used to compute the final output embedding $y$. The converse strategy shown in Figure~\ref{fig:routed-ffn}b, i.e., pruning the columns of $W_I$ and rows of $W_O$, does not work because it means that some of the input dimensions are not used to compute the intermediate output $h$ and some dimensions of the final output $y$ are zero. This severely reduces model capacity and thus degrades model quality.

The final problem becomes how to choose the blocks to prune. We cannot compute the exact output $y$ and then select the smallest output dimensions because this does not save computation.
Thus, we train a small \textit{route network} to guide pruning. Specifically, assuming that $W_I$ has $D$ rows, we organize the rows into $G$ groups by assigning $D/G$ adjacent rows to each group.
The route network is a single layer feed-forward network $x_R=xW_R$ with $W_R\in \mathbb{R}^{d\times G}$, where $d$ is the embedding dimension; and we set the top $G'$ entries in $x_R$ with the largest magnitude (and the corresponding row groups in $W_I$) as activated.
Note that the activated groups of $W_O$ are decided according to the activated groups of $W_I$, and thus we do not need another route network for $W_O$. We use a small $G$ (e.g., 4 or 8) such that the route network is cheap to compute.
The route network is trained along with the FFN such that it can learn to determine the weight groups to activate. We introduce a load-balancing loss for the route network to ensure that the weight groups have similar activation rates and that the model capacity is fully utilized.

%\yt{(TODO: fix citations)}

% Optimized Runtime
\section{Sparse Execution Engine}
\label{sec:system}

To implement \name{}, we utilize PyTorch's existing modules such as embedding layers, fully connected layers, and back-propagation training.
However, PyTorch does not support the sparse MHA and routed FFN modules we proposed in Section~\ref{sec:algorithm}.
Thus, we customize CUDA implementations for the two modules and discuss the key optimizations in this part.

\subsection{Sparse MHA}\label{subsec:sys-mha}

Recall that our sparse MHA takes 3 steps, i.e., quantizing the query and key vectors using PQ codebooks, finding the top-$L$ query-key pairs based on PQ quantization, and finally computing sparse attention for the top-$L$ query-key pairs.
We customize the 3 steps on GPU to make sparse MHA efficient.

\begin{algorithm}[!t]
    \caption{The procedure of PQ quantization.}
    \label{alg:pq-quantization}
    \begin{flushleft}
        \textbf{Input:} the vectors to quantize $X$, and the codebooks $\mathcal{C}$\\
        \textbf{Output:} the codewords $C_X$ and quantization errors $e_X$ for $X$
    \end{flushleft}
    \begin{algorithmic}[1]
        \For{subspace from $i=1$ \textbf{to} $M$}
        \State Compute distances $Dist_i \gets \mathsf{cdist}(X_i,\ \mathcal{C}_i )$
        \State Get centroids $C_X[i] \gets \mathsf{argmin}( Dist_i,\ dim=-1 )$
        \State Select codewords $Code_X[i] \gets \mathsf{index\_select}( \mathcal{C}_i,\ C_X[i])$
        \State Compute errors $e_X \gets \mathsf{DKM}(Code_X[i],\ X_i )$
        \EndFor
    \end{algorithmic}
\end{algorithm}

\stitle{PQ quantization}
We use Algorithm~\ref{alg:pq-quantization} to quantize the query and key vectors, and each vector is mapped to its nearest codewords in the $M$ codebooks.
The vectors of a sequence are processed together for one codebook at a time, and thus the input dimension of the distance computation operator (\textsf{cdist}) is \textsf{[sequence length, dim codeword]}, which contains a subspace of all the input vectors; the dimension of a codebook is \textsf{[num codewords, dim codeword]}, and the distance output is of size \textsf{[sequence length, num codewords]}.
The \textsf{argmin} operator finds the nearest codeword for each input vector in the codebook, and the output dimension is \textsf{[sequence length, 1]}.
We utilize the differentiable k-means (DKM)~\cite{cho22dkm} algorithm to adjust the codebook, which allows the codebook to match data distribution and requires the computation of the quantization error (i.e., the distance between a vector and its nearest codeword).
%This allows the codebook to match data distribution, resulting in a compact codebook that minimizes the quantization error between the input data and the nearest codeword.
%\yt{(the benefit of 16 on GPU is removed, because use 32 or 64 is also efficient.)}

We set the codeword dimension of each codebook (i.e., $d'$) to 8 and codebook size (i.e., the number of codewords in each codebook) to 16.
Using more codewords improves quantization accuracy but also increases computation cost. We observe that 16 codewords already yield good model quality as the softmax output of attention tends to converge to a small number of high probability values. We accelerate Algorithm~\ref{alg:pq-quantization} with two implementation optimizations.
First, we perform codebook update every 20 mini-batches instead of every mini-batch, which means that Lines 4-5 of Algorithm~\ref{alg:pq-quantization} are skipped to reduce computation for most mini-batches.
This does not harm model quality because the codebooks represent centroids of the query and key vectors, which change slowly over mini-batches.
Second, the \textsf{cdist} and \textsf{argmin} operators are fused into one CUDA kernel.
This avoids producing the intermediate results for \textsf{cdist}, whose size is large (i.e., \textsf{[sequence length, num codewords]}), and thus reduces memory access.

\begin{algorithm}[t!]
    \caption{The procedure of top-$L$ selection.}
    \label{alg:topk-lookup}
    \begin{flushleft}
        \textbf{Input:} the codewords $C_Q, C_K$ for the query and key vectors \\
        \textbf{Output:} the position of top-$L$ query-key pairs as $Indices$
    \end{flushleft}
    \begin{algorithmic}[1]
        \ForAll{query codeword $c_q$ \textbf{in} $C_Q$}
        \State $Ptr \gets \mathsf{allocate}(M + 1)$, $Bucket \gets \mathsf{allocate}((M + 1) \times L)$
        \ForAll{key codeword $c_k$ \textbf{in} $C_K$}
        \State  $s \gets \mathsf{indicator}(c_q,\ c_k)$ \Comment{Distance with PQ code}
        \State $ptr \gets Ptr[s]$ \Comment{Position to store the key}
        \State $Bucket[s][ptr] \gets$ the index of key
        \State $Ptr[s] \gets \mathsf{min}(ptr + 1,\ L - 1)$ \Comment{Prevent overflow}
        \EndFor
        \State $s \gets M$, $ptr \gets 0$ \Comment{Pointer used to read from $Bucket$}
        \ForAll{$i=0$ \textbf{to} L - 1}
        \If{$ptr$ \textbf{equals} $Ptr[s]$} \Comment{$ptr$ reaches the end}
        \State $ptr \gets 0$,\ $s \gets s - 1$ \Comment{Move to the previous bucket}
        \EndIf
        \State $Indices_q[i] \gets Bucket[s][ptr++]$ \Comment{Write result for $q$}
        \EndFor
        \EndFor
    \end{algorithmic}
\end{algorithm}

\stitle{Top-$L$ selection}
For the query and key vectors of a sequence, we use Algorithm~\ref{alg:topk-lookup} to find the top-$L$ most similar key vectors for each query vector. The procedure works as follows.

\squishlist
\item \textit{Initialize (line 2).}
As each query and key vector is quantized to $M$ codewords, we use the number of common codewords (i.e., \textsf{indicator} in line 4) to measure their similarity.
Since the possible values of common codewords fall in $\{0,\ M\}$, we allocate $M+1$ empty buckets, indexed from $Bucket[0]$ to $Bucket[M]$, to hold the keys with the corresponding similarity scores for each query vector.
We set the capacity of each bucket as $L$ such that each bucket can hold all the top-$L$ keys in the worst case.

\item \textit{Assign (lines 3-8).}
The algorithm goes over the keys to compute the similarity scores.
For a key with score $s$, it is put into $Bucket[s]$ (line 6); and we use $Ptr[s]$ to record the next position to insert for $Bucket[s]$. If a bucket already contains $L$ keys, we overwrite an old key with the new key to avoid bucket overflow (line 7). This is because only $L$ top keys are required and dynamically allocating memory on GPU is impossible.
At this time, the algorithm groups keys together, enabling top-$L$ selection from the buckets.

\item \textit{Retrieve (lines 9-16).}
We retrieve the top-$L$ similar keys for each query vector by checking the buckets from high index to low index (i.e., $Bucket[M]$ to $Bucket[0]$). This is because the keys in higher buckets have larger similarity scores w.r.t. the query vector. This process stops when $L$ keys are collected.
Note that we do not need to perform a full sort within each bucket as in classical bucket sort.
\squishend

%, which avoids the computational cost of sorting. Yet it is still able to obtain the top-$L$ most similar keys
Algorithm~\ref{alg:topk-lookup} is efficient for GPU execution.
The outer loop (line 1) can be easily parallelized by GPU threads, i.e., each query vector is processed by one GPU thread in parallel.
We also parallelize the inner loops (lines 3 and 10) by scheduling two extra GPU threads to process different bucket elements, i.e., one thread handles odd indices and the other handles even indices.
% \yt{Note that there is no race condition because two threads are accessing and modifying completely separate bucket elements in the inner loops.} \xiao{BTW, to collect L keys, line 10 needs atomic access to the number of collected keys, also line 3 needs to update the next position when inserting each key}
Moreover, we allocate the buckets in fast on-chip shared memory and only move the final top-$L$ query-key pairs to global memory when collecting the results. This requires only one sequential write and has a small memory access overhead.
Last but not least, our bucket-sort-based top-$L$ selection is much more efficient than using the standard approximate distance of PQ, which computes and sorts floating point distances.

%as they store only non-zero values in the matrices and thus yield memory savings
\stitle{Sparse attention computation}
We utilize sparse operations to compute the sparse attention matrix and the attention outputs.
In particular, sampled dense-dense matrix multiplication (SDDMM) is used to compute the attention weights between the top-$L$ query and key pairs.
Sparse matrix-matrix multiplication (SpMM) is used to multiply the sparse attention weights with the value vectors to generate the attention outputs.
Instead of implementing SDDMM and SpMM from scratch, we adapt NVIDIA's cuSPARSE library.

The cuSPARSE library allows storing sparse matrices using either the compressed sparse row (CSR) format or coordinate list (COO) format, which has different performance characteristics.
In particular, CSR stores row pointers (\textit{Indptr}) and column indices (\textit{Indices}), and thus is efficient for row-major arithmetic operations like matrix multiplication. COO stores the row and column coordinates and makes it simple to construct a sparse matrix.
We choose the CSR format for the sparse matrices because it is efficient for row-oriented attention computation.

\begin{figure}[!t]
    \centering
    \includegraphics[width=0.80\columnwidth]{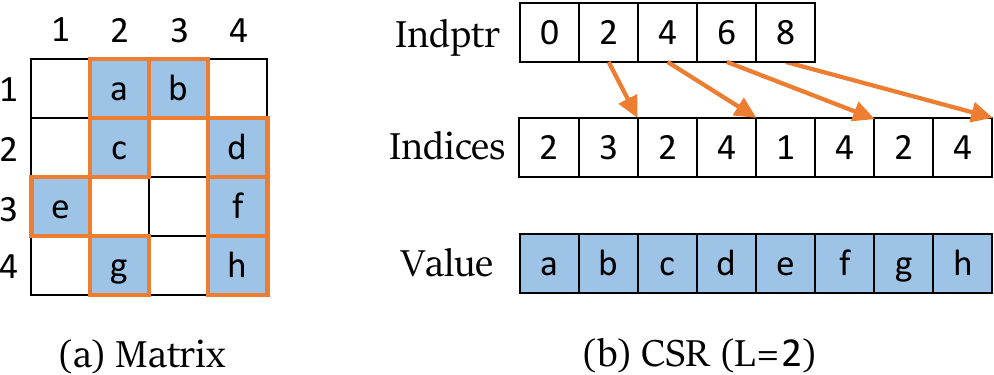}
    \caption{CSR encoding from the top-$L$ query-key pairs, the non-zero entries are colored.}
    \label{fig:csr-encoding}
\end{figure}

The CSR matrix used for SDDMM and SpMM is constructed directly from the output of the previous top-$L$ selection step. As shown in Figure~\ref{fig:csr-encoding}, the \textit{Indices} array of CSR stores the column indices of the non-zero values (i.e., top-$L$ keys), which is the output of the top-$L$ selection step. The \textit{Indptr} array marks the start and end positions of the \textit{Indices} for each row. Since we select $L$ keys for each query vector, the \textit{Indptr} array is $[0, L, 2L, 3L, ...]$. Note that we only need to construct the CSR matrix once and reuse it for both SDDMM and SpMM, this is because the sparse matrix produced by SDDMM (i.e., the attention matrix) serves as the input to SpMM (i.e., for final attention output), and the \textit{softmax} function does not change the matrix shape.

\subsection{Blocked Sparse Matrix-Vector Multiply}\label{subsec:sys-ffn}

\begin{algorithm}[t!]
    \caption{The procedure of BSpMV.}
    \label{alg:block-ffn}
    \begin{flushleft}
        \textbf{Input:} token sequence $X$, model weight matrices $W_I$, $W_O$, and $Indices$ of the activated model weight blocks for the tokens \\
        \textbf{Output:} a new sequence representation $Y$
    \end{flushleft}
    \begin{algorithmic}[1]
        \For{block from $i=1$ \textbf{to} $G$}
        \State Generate token mask $Mask_T \gets \mathsf{eq}(Indices,\ i)$
        \State Select tokens $X_i \gets X[Mask_T]$
        \State Compute inner projection $H \gets \mathsf{ReLU}(X_i W_I[i])$
        \State Compute outer projection $Y[Mask_T] \gets H W_O[i]$
        \EndFor
    \end{algorithmic}
\end{algorithm}

The standard dense FFN layer takes the token sequences as input and performs batched matrix multiplications between the input tokens and the weight matrices. Our routed FFN layer activates different subsets of the weight matrices for each input token, i.e., different tokens interact with different weight matrices.
This leads to a general sparse-dense matrix-vector multiplication problem that invalidates batched matrix multiplications and harms efficiency.

A naive solution is to generate a binary mask matrix that indicates the activated weights for each input token, and these masks are multiplied with the weight matrices in an element-wise manner before batch computation. This solution has high memory consumption because it stores a separate mask for each input token.
To reduce memory consumption, one may use the block-compressed sparse row (BSR) matrix format to store the masks. Like CSR, BSR stores only the nonzero entries but each entry corresponds to a sub-matrix instead of scalar values. For a sequence length $n$ and number of blocks $\hat{B}$, BSR requires $O(n\hat{B})$ space.
BSR is still inefficient because duplicating the weight matrices for each token has a high overhead.

We propose an efficient blocked sparse matrix-vector multiplication (BSpMV) method to perform batched multiplications between sparse FFN matrices and vectors.
The insight is that the sparsity induced by our routed FFN actually means that each dense block of weights is only relevant for computing the outputs for a subset of the input tokens.
In particular, as shown in Algorithm~\ref{alg:block-ffn}, we iterate over the weight blocks (line 1) and select a subset of token vectors (line 3) to perform dense matrix computation with the block (lines 4 and 5).
Our BSpMV approach has several benefits: (1) it can leverage fast dense GEMM computations for each block; (2) the processing of different weight blocks can be parallelized across GPU streams; (3) the number of blocks is small, and thus it is fast to loop over them; (4) each block in the iteration can compute some output results without synchronizing with the other blocks.
BSpMV needs to extract the relevant token vectors for each block but the overhead is small.

% Evaluation
\section{Experimental Evaluation}\label{sec:eval}

In this part, we evaluate the efficiency and effectiveness of \name{} with extensive experiments. The key findings are that:
\squishlist
\item \name{} outperforms both full-parameter tuning and LoRA fine-tuning across all model configurations, with a maximum running time speedup of $2.2\times$ and peak memory reduction of 50\%.
\item Our two key optimizations (i.e., sparse MHA and routed FFN) and their implementations are effective in improving efficiency.
\squishend

\subsection{Experiment Settings}\label{sec:eval-setting}

\begin{table}[!t]
  \small
  \centering
  \setlength{\tabcolsep}{1.75mm}
  \caption{Statistics of the Transformer blocks in experiments.}
  \label{tab:eval-workload}
  \begin{tabular}{cccccc}
    \toprule
    \textbf{Name} & $d_{model}$ & $d_{head}$ & $d_{ffn}$ & \textbf{Pre-trained Model} \\
    \midrule
    OPT-1024      & 1024        & 64         & 4096      & GPT2-medium, OPT-350M      \\
    OPT-2048      & 2048        & 64         & 8192      & OPT-1.3B                   \\
    OPT-2560      & 2560        & 80         & 10240     & OPT-2.7B                   \\
    LLaMA-2560    & 2560        & 128        & 6912      & Sheared-LLaMA-2.7B         \\
    LLaMA-4096    & 4096        & 128        & 11008     & Open-LLaMA-7B              \\
    \bottomrule
  \end{tabular}
\end{table}

\stitle{Models}
We use two well-known Transformer-based large language models to evaluate the performance of end-to-end fine-tuning, i.e., OPT-2.7B~\cite{zhang2022opt} and LLaMA-2.7B~\cite{xia2023sheared}.
In particular, OPT-2.7B stacks 32 Transformer blocks, and uses the ReLU activation function and a simple feed-forward structure. LLaMA-2.7B also stacks 32 Transformer blocks but uses the GeLU activation function and rotary positional embeddings~\cite{su21rotary}. We also experiment with the 5 Transformer configurations in Table~\ref{tab:eval-workload} to evaluate the performance of individual Transformer blocks.
In particular, $d_{model}$ is the embedding dimension size of the Transformer models (including multi-heads), $d_{head}$ represents the token embedding size of a single attention head, which corresponds to $d$ in \S\ref{subsec:bg-transformers}, and $d_{ffn}$ means the hidden dimension size in FFN.
These Transformer blocks come from popular language models listed in the last column of Table~\ref{tab:eval-workload} and thus are representative of Transformer configurations.

\stitle{Datasets} We use the following 3 datasets for the experiments.
\squishlist
\item \textit{Massive Multi-task Language Understanding} (MMLU)~\cite{hendrycks21mmlu} is a popular dataset for question answering, where 4 choices are provided for each question and the model is trained to select an answer for each question. It uses a metric also called MMLU to evaluate the question answering accuracy of the models, and a higher MMLU refers to better model quality.
\item \textit{Wikitext-103}~\cite{merity2016wikitext} is a widely used dataset for next-word prediction training. Collected from Wikipedia, it contains over 100 million tokens. Wikitext-103 is an unlabeled dataset and thus we use perplexity (PPL) as an evaluation metric to assess the model quality. A smaller PPL score indicates that the language model fits the training data better and has lower losses.
\item \textit{Random} contains randomly generated sequences of arbitrary length that can be specified as needed. We mainly use it for micro experiments due to this flexibility.
\squishend

\stitle{Baselines and platform}
We compare the \name{} system with full-parameter tuning (i.e., \textit{Full}) and LoRA fine-tuning (i.e., \textit{LoRA}), both of which have been well implemented and extensively optimized. By default, the sparse MHA of \name{} uses the top-$1/8$ attention weights for each query and the routed FFN activates $1/2$ of the parameters in the projection matrices. Our experiment platform is a server with 4 NVIDIA RTX3090 GPUs (each with 24GB global memory), dual Intel Xeon Gold 6326 CPU, and 256 GB main memory. All the experiments are conducted with single-precision floating point numbers, and weight decay is enabled for the optimizer.

\subsection{Main Results}
\label{sec:eval-overview}

\begin{table}[!t]
      \small
      \centering
      \setlength{\tabcolsep}{1.25mm}
      \caption{End-to-end fine-tuning results. MMLU indicates model quality, and the speedups are over Full.}
      \begin{tabular}{ccccc}
            \toprule
            \textbf{Model} & \textbf{System} & \textbf{MMLU} & \textbf{Max Length} & \textbf{Time (speedup)} \\
            \midrule
            \multirow{3}{*}{OPT-2.7B}
                           & Full            & $27.0$        & 256                 & 6.7 h (1.00x)           \\
                           & LoRA            & $27.0$        & 512                 & 5.8 h (1.15x)           \\
                           & \name{}         & $26.1$        & 768                 & 4.6 h (1.47x)           \\
            \midrule
            \multirow{3}{*}{LLaMA-2.7B}
                           & Full            & $29.0$        & 256                 & 6.9 h (1.00x)           \\
                           & LoRA            & $29.0$        & 384                 & 5.8 h (1.18x)           \\
                           & \name{}         & $28.4$        & 640                 & 5.0 h (1.39x)           \\
            \bottomrule
      \end{tabular}
      \label{tab:eval-overview}
\end{table}

%\begin{table}[!t]
%	\small
%	\centering
%	\setlength{\tabcolsep}{1.0mm}
%	\caption{End-to-end fine-tuning results.}
%	%
%	\begin{tabular}{cccccc}
%		\toprule
%		Model & Method  & Parameters & MMLU           & Max Length & Speedup \\
%		\midrule
%		\multirow{3}{*}{OPT-2.7B}
%		& Full    & 2651.49 M  & $32.8 \pm 2.3$ & 256        & 1.0x    \\
%		& LoRA    & 24.50 M    & $32.8 \pm 0.4$ & 512        & 1.15x   \\
%		& \name{} & 24.54 M    & $31.7 \pm 0.4$ & 768        & 1.47x   \\
%		\midrule
%		\multirow{3}{*}{LLaMA-2.7B}
%		& Full    & 2576.41 M  & $33.9 \pm 1.9$ & 256        & 1.0x    \\
%		& LoRA    & 25.09 M    & $33.8 \pm 0.3$ & 384        & 1.18x   \\
%		& \name{} & 25.15 M    & $32.9 \pm 0.3$ & 640        & 1.39x   \\
%		\bottomrule
%	\end{tabular}
%	\label{tab:eval-overview}
%\end{table}

\stitle{End-to-end fine-tuning}
In this experiment, we conduct end-to-end fine-tuning for the OPT-2.7B and LLaMA-2.7B models on the MMLU dataset.
To emulate real-world fine-tuning scenarios, we employ all 4 GPUs on our system and use DeepSpeed~\cite{rasley20deepspeed} for multi-GPU capabilities. Note that DeepSpeed has a sophisticated memory management mechanism for multi-GPU configuration, and we enable parameter and activation offloading in DeepSpeed from GPU memory to CPU memory to reduce GPU memory footprint. As a result, we are unable to measure the peak memory consumption on GPUs.
To understand the memory constraints of the baselines, we utilize the maximum sequence length (Max Length) without triggering an out-of-memory (OOM) error as a surrogate. DeepSpeed produces an OOM error when a single Transformer block cannot fit into GPU memory. We adjust the sequence length in increments of 128 until this OOM state occurs.
Note that to ensure a fair comparison, we run all baselines with a sequence length of 512 and measure the 5-shot learning capabilities of MMLU. Each model is trained for 10k mini-batches with a batch size of 16.

The results in Table~\ref{tab:eval-overview} show that \name{} reduces running time and increases maximum sequence length compared with both full-parameter tuning and LoRA. Regarding running time for the two models, \name{} speeds up full-parameter tuning by 1.47x and 1.39x. \name{} is also 1.28x and 1.18x faster than LoRA. This is remarkable because the speedup of \name{} over LoRA is larger than LoRA over full-parameter tuning.
Moreover, \name{} supports 2x sequence length compared with full-parameter tuning and over 1.5x compared with LoRA, which indicates a significant memory saving.
Considering model quality, a small model quality degradation is observed in \name{} in terms of MMLU decrease.
Thus, \name{} achieves a good trade-off between efficiency and model quality.

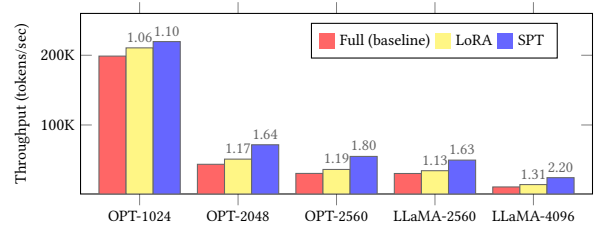
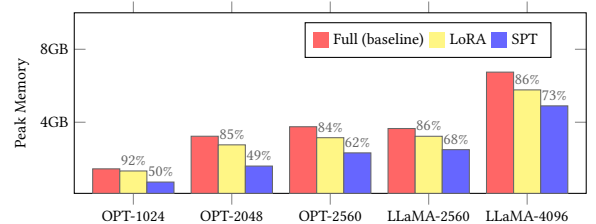
\begin{figure}[t!]
    \centering
    \scriptsize
    % throughput
    \begin{subfigure}[t]{0.45\textwidth}
        \begin{tikzpicture}
            \begin{axis}[
                    axis lines=box,
                    scale only axis=true,
                    ybar,
                    ymin=1.0,
                    ymax=260.0,
                    width=0.85\textwidth,
                    height=0.30\textwidth,
                    ylabel=Throughput (tokens/sec),
                    xmin={[normalized]-0.6},
                    xmax={[normalized]4.6},
                    xtick distance=1.0,
                    yticklabel={$\pgfmathprintnumber{\tick}$K},
                    symbolic x coords={
                            OPT-1024, OPT-2048, OPT-2560, LLaMA-2560, LLaMA-4096
                        },
                    point meta=explicit symbolic,
                    nodes near coords,
                    nodes near coords align={horizontal},
                    bar width=0.045\textwidth,
                    legend cell align={center},
                    legend image code/.code={%
                            \draw[#1, draw=none] (0cm,-0.1cm) rectangle (0.2cm,0.1cm);
                        },
                    legend style={anchor=south west, at={(0.45,0.75)}, legend columns=3}
                ]
                % full
                \addplot+ [black!60,fill=red!60,bar shift=-0.045\textwidth]
                table[x=x, y=y, row sep=\\] {
                        x           y     \\
                        OPT-1024         198.8  \\
                        OPT-2048         43.6  \\
                        OPT-2560        30.5  \\
                        LLaMA-2560        30.4  \\
                        LLaMA-4096        11.1  \\
                    };
                % lora, postaction={pattern=dots}
                \addplot+ [black!60,fill=yellow!60,bar shift=0.00\textwidth]
                table[x=x, y=y, meta=meta, row sep=\\] {
                        x           y               meta      \\
                        OPT-1024         210.6      1.06      \\
                        OPT-2048         51.0       1.17      \\
                        OPT-2560        36.2        1.19      \\
                        LLaMA-2560        34.4      1.13      \\
                        LLaMA-4096        14.5      1.31      \\
                    };
                % sparse, postaction={pattern=north east lines}
                \addplot+ [black!60,fill=blue!60,bar shift=0.045\textwidth]
                table[x=x, y=y, meta=meta, row sep=\\] {
                        x           y               meta       \\
                        OPT-1024         219.6      1.10      \\
                        OPT-2048         71.6       1.64      \\
                        OPT-2560        54.9        1.80      \\
                        LLaMA-2560        49.6      1.63      \\
                        LLaMA-4096        24.4      2.20      \\
                    };
                \legend{Full (baseline), \space\space LoRA, \space\space \name{}}
            \end{axis}
        \end{tikzpicture}
        \caption{Training throughput for different Transformer blocks, the speedups are over full-parameter tuning.}
    \end{subfigure}
    % peak memory
    \begin{subfigure}[t]{0.45\textwidth}
        \begin{tikzpicture}
            \begin{axis}[
                    axis lines=box,
                    scale only axis=true,
                    ybar,
                    ymin=0.1,
                    ymax=10.0,
                    ytick distance=4.0,
                    width=0.85\textwidth,
                    height=0.30\textwidth,
                    ylabel=Peak Memory,
                    xmin={[normalized]-0.6},
                    xmax={[normalized]4.6},
                    xtick distance=1.0,
                    yticklabel={$\pgfmathprintnumber{\tick}$GB},
                    symbolic x coords={
                            OPT-1024, OPT-2048, OPT-2560, LLaMA-2560, LLaMA-4096
                        },
                    point meta=explicit symbolic,
                    nodes near coords,
                    nodes near coords align={horizontal},
                    bar width=0.045\textwidth,
                    legend cell align={center},
                    legend image code/.code={%
                            \draw[#1, draw=none] (0cm,-0.1cm) rectangle (0.2cm,0.1cm);
                        },
                    legend style={anchor=south west, at={(0.45,0.75)}, legend columns=3}
                ]
                % full
                \addplot+ [black!60,fill=red!60,bar shift=-0.045\textwidth]
                table[x=x, y=y, row sep=\\] {
                        x             y      \\
                        OPT-1024      1.444  \\
                        OPT-2048      3.232  \\
                        OPT-2560      3.753  \\
                        LLaMA-2560    3.655  \\
                        LLaMA-4096    6.747  \\
                    };
                % lora
                \addplot+ [black!60,fill=yellow!60,bar shift=0.00\textwidth]
                table[x=x, y=y, meta=meta, row sep=\\] {
                        x             y         meta       \\
                        OPT-1024      1.328     92\%       \\
                        OPT-2048      2.760     85\%       \\
                        OPT-2560      3.152     84\%       \\
                        LLaMA-2560    3.230     86\%       \\
                        LLaMA-4096    5.773     86\%       \\
                    };
                % sparse
                \addplot+ [black!60,fill=blue!60,bar shift=0.045\textwidth]
                table[x=x, y=y, meta=meta, row sep=\\] {
                        x           y             meta     \\
                        OPT-1024         0.720    50\%     \\
                        OPT-2048         1.597    49\%     \\
                        OPT-2560         2.320    62\%     \\
                        LLaMA-2560       2.496    68\%     \\
                        LLaMA-4096       4.897    73\%     \\
                    };
                \legend{Full (baseline), \space\space LoRA, \space\space \name{}}
            \end{axis}
        \end{tikzpicture}
        \caption{Peak memory consumption for training the 5 Transformer blocks, the percentages are over full-parameter tuning.}
    \end{subfigure}
    \caption{Fine-tuning throughput and peak memory for 5 Transformer blocks (batch size 16, sequence length 512).}
    \label{fig:eval-overall}
\end{figure}

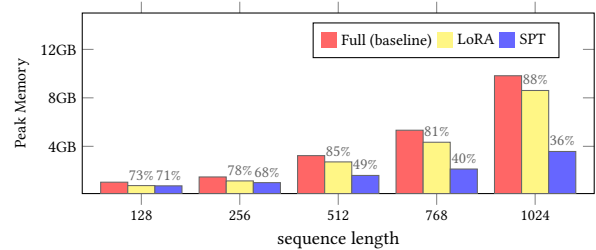
\begin{figure}[t!]
    \centering
    \scriptsize
    % seq_length
    \begin{subfigure}[t]{0.45\textwidth}
        \begin{tikzpicture}
            \begin{axis}[
                    axis lines=box,
                    scale only axis=true,
                    ybar,
                    ymin=0.1,
                    ymax=15.0,
                    % ymode=log,
                    % log ticks with fixed point,
                    ytick distance=4.0,
                    width=0.85\textwidth,
                    height=0.30\textwidth,
                    ylabel=Peak Memory,
                    xmin={[normalized]-0.6},
                    xmax={[normalized]4.6},
                    xtick distance=1.0,
                    xlabel={\footnotesize{sequence length}},
                    yticklabel={$\pgfmathprintnumber{\tick}$GB},
                    symbolic x coords={
                            128, 256, 512, 768, 1024
                        },
                    point meta=explicit symbolic,
                    nodes near coords,
                    nodes near coords align={horizontal},
                    bar width=0.045\textwidth,
                    legend cell align={center},
                    legend image code/.code={%
                            \draw[#1, draw=none] (0cm,-0.1cm) rectangle (0.2cm,0.1cm);
                        },
                    legend style={anchor=south west, at={(0.45,0.75)}, legend columns=3}
                ]
                % full
                \addplot+ [black!60,fill=red!60,bar shift=-0.045\textwidth]
                table[x=x, y=y, row sep=\\] {
                        x           y     \\
                        128         1.040  \\
                        256         1.472  \\
                        512         3.232  \\
                        768         5.329  \\
                        1024        9.824  \\
                    };
                % lora
                \addplot+ [black!60,fill=yellow!60,bar shift=0.00\textwidth]
                table[x=x, y=y, meta=meta, row sep=\\] {
                        x           y       meta     \\
                        128         0.759   73\%     \\
                        256         1.145   78\%     \\
                        512         2.714   85\%     \\
                        768         4.336   81\%     \\
                        1024        8.607   88\%     \\
                    };
                % sparse
                \addplot+ [black!60,fill=blue!60,bar shift=0.045\textwidth]
                table[x=x, y=y, meta=meta, row sep=\\] {
                        x           y        meta  \\
                        128         0.738     71\%     \\
                        256         1.003     68\%     \\
                        512         1.597     49\%   \\
                        768         2.126     40\%   \\
                        1024        3.574     36\%   \\
                    };
                \legend{Full (baseline), \space\space LoRA, \space\space \name{}}
            \end{axis}
        \end{tikzpicture}
    \end{subfigure}
    \caption{Peak memory requirements when using different sequence lengths (OPT-2048, batch size 16).}
    \label{fig:eval-sequence}
\end{figure}

\stitle{Transformer blocks}
To understand the performance gains over individual Transformer blocks, we evaluate \name{} on the 5 different Transformer configurations listed in Table~\ref{tab:eval-workload}. We use randomly generated sequences to easily control the experiment settings and run the evaluations on a single GPU to eliminate the overhead of multi-GPU communication.
We transform the running time, i.e., time to compute the forward and backward passes for a Transformer block, into throughput (i.e., tokens processed per second) for more intuitive comprehension of the speedups. All results are averaged over 20 runs.

As reported in Figure~\ref{fig:eval-overall}, \name{} yields consistent benefits in terms of throughput speedup (i.e., shorter training time) and lower peak memory requirements than full-parameter tuning and LoRA for all the configurations. Specifically, \name{} achieves higher throughput than full-parameter tuning, with speedups ranging from 1.10x to 2.20x. It also outperforms LoRA, with throughput improvements of 1.04x to 1.68x.
The maximum speedup is observed on LLaMA-4096. This is because LLaMA-4096 has the largest FFN dimension (cf. Table~\ref{tab:eval-workload}) among the given Transformer settings, where FFN becomes dominant and exposes more optimization opportunities for the routed FFN module.
On the other hand, \name{} has much lower peak memory usage than the other systems. Specifically, it uses only 50\% to 73\% peak memory compared with full-parameter tuning, and 54\% to 85\% with LoRA.
The largest peak memory reduction is observed on OPT-1024. This is because OPT-1024 has the smallest FFN dimension while MHA dominates most of the peak memory requirements, and thus the memory saving of our sparse MHA is significant.
We also observed that varying both the sequence length and batch size did not impact the speedup provided by the \name{}. This is mainly due to the routed FFN treating all additional dimensions as batches of tokens. The speedup stays consistent regardless of sequence length or batch size changes.
Note that the speedups of \name{} shown in Figure~\ref{fig:eval-overall} are generally higher than those in Table~\ref{tab:eval-overview}. This is because factors like cross-GPU communication slowdowns the end-to-end tuning, where these costs are not reduced by our optimizations.

Figure~\ref{fig:eval-sequence} shows that the peak memory consumption rises rapidly as the sequence length grows. This is because the attention matrix grows quadratically with sequence length.
LoRA reduces memory consumption compared with full-parameter tuning, but the effect is limited. In comparison, \name{} achieves more substantial memory savings for longer sequences as a result of MHA becoming more predominant.
We also varied the batch size and found that it had minimal impact on the memory performance, since the MHA operates along the sequence dimension. Increasing the batch size does not affect the mechanism itself.

\stitle{Discussions}
LoRA states to reduce hardware requirements (especially memory) by using much smaller trainable parameters~\cite{hu2022lora}. However, the memory savings of LoRA are not obvious in our experiments, e.g., taking up 84\% to 92\% of full-parameters tuning in Figure~\ref{fig:eval-overall}, this is because we use a relatively large batch size, and intermediate activations (rather than model parameters) dominate memory consumption in the case.
For example, consider a fully connected (FC) layer with dimension $4096 \times 1024$, the model parameters take $4M$, which can be saved by LoRA.
However, with a batch size of 16 and sequence length of 512, the activations take up $16 \times 512 \times 4096 = 32M$, which are much larger than the parameters and can not be saved by LoRA. In contrast, the sparse MHA of \name{} reduces the attention weights for each sequence, and thus consistent memory reduction is achieved at any batch size.
Moreover, \name{} has limited speedups when fine-tuning under extreme conditions, e.g., a batch size of 1, sequence length of 64, and FFN hidden dimension of 128. This is because when the total number of tokens input to the model is small, the optimization opportunities for routed FFN are limited.
However, these cases are rare in practice, and the trends are to use longer sequences to capture richer contexts and larger FFNs to increase model capacity. Thus, \name{} may yield even more significant efficiency gains for larger models that will come in the future.

\subsection{Parameter Sensitivity and Design Choices}
\label{sec:eval-ablation}

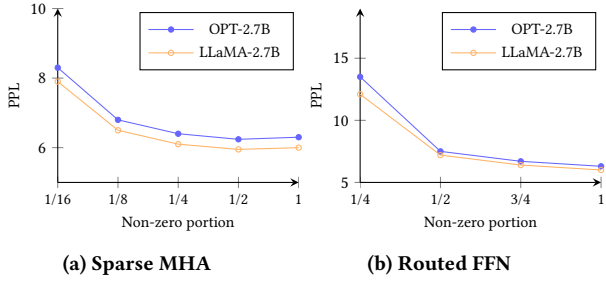
\begin{figure}[!t]
    \centering
    \scriptsize
    \begin{subfigure}[t]{0.2\textwidth}
        \begin{tikzpicture}
            \begin{axis}[
                    axis lines=left,
                    xlabel=Non-zero portion,
                    ylabel=PPL,
                    ymin=5.0,
                    ymax=10.0,
                    symbolic x coords={1/16, 1/8, 1/4, 1/2, 1},
                    scale only axis=true,
                    width=0.9\textwidth,
                    height=0.65\textwidth,
                    legend cell align={center},
                ]
                \addplot[blue!60, mark=*, mark size=1pt] coordinates {
                        (1/16,8.3) (1/8,6.8) (1/4,6.4) (1/2,6.24) (1,6.3)
                    };
                \addplot[orange!60, mark=o, mark size=1pt] coordinates {
                        (1/16,7.9) (1/8,6.5) (1/4,6.1) (1/2,5.95) (1,6.0)
                    };
                \legend{OPT-2.7B, LLaMA-2.7B}
            \end{axis}
        \end{tikzpicture}
        \caption{Sparse MHA}
    \end{subfigure}
    \hspace{0.02\textwidth}
    \begin{subfigure}[t]{0.2\textwidth}
        \begin{tikzpicture}
            \begin{axis}[
                    axis lines=left,
                    xlabel=Non-zero portion,
                    ylabel=PPL,
                    ymin=5.0,
                    ymax=19.0,
                    symbolic x coords={1/4, 1/2, 3/4, 1},
                    scale only axis=true,
                    width=0.9\textwidth,
                    height=0.65\textwidth,
                    legend cell align={center},
                ]
                %
                % \addplot[black!60,dotted,style=very thick] coordinates {
                %         (1/4,12.0) (1,12.0)
                %     };
                \addplot[blue!60, mark=*, mark size=1pt] coordinates {
                        (1/4,13.5) (1/2,7.5) (3/4,6.7) (1,6.3)
                    };
                \addplot[orange!60, mark=o, mark size=1pt] coordinates {
                        (1/4,12.1) (1/2,7.2) (3/4,6.4) (1,6.0)
                    };
                \legend{OPT-2.7B, LLaMA-2.7B}
            \end{axis}
        \end{tikzpicture}
        \caption{Routed FFN}
    \end{subfigure}
    \caption{The influence of sparsity on model quality.}
    \label{fig:eval-quality}
\end{figure}

\stitle{Model quality}
In this experiment, We test the quality impact of \name{} under different sparsity strengths. In particular, the sparsity of MHA is specified by the portion of non-zero attention weights, while the sparsity of routed FFN is determined by the portion of activated model parameters.
We run this experiment using the Wikitext-103 dataset and evaluate PPL to measure model quality. Note that PPL is a strict measurement as it exponentiates the loss. Therefore, small differences in loss can result in large differences in PPL.
The increasing PPLs shown in Figure~\ref{fig:eval-quality} stabilize when the non-zero portion reaches 1/8 for sparse MHA and 1/2 for routed FFN. This indicates that our default sparsity strengths achieve a good balance between quality and efficiency.
Here MHA allows higher sparsity than FFN because (1) there are multiple attention heads, which can complement each other to combat the information loss caused by sparsifying, and (2) the attention weights produced by the softmax function are more skewed than the FFN projection as we have shown in \S~\ref{sec:algorithm}, thus using sparse MHA yields less information loss.

\begin{table}[!t]
       \small
       \centering
       \setlength{\tabcolsep}{2.5mm}
       \caption{The running time and peak memory consumption of MHA and FFN when using different sparsity.}
       \label{tab:eval-profile}
       % 109 mm, 161 mm
       \subcaption{OPT-2048}
       \begin{tabular}{cccc}
              \toprule
              Module & Method          & Peak Mem & Duration \\
              \midrule
              \multirow{4}{*}{MHA}
                     & LoRA            & 2626 MB  & 52.5 ms  \\
                     & LoRA (compiled) & 2418 MB  & 47.1 ms  \\
                     & SPT (1/4)       & 1784 MB  & 72.9 ms  \\
                     & SPT (1/8)       & 1123 MB  & 54.1 ms  \\
              \midrule
              \multirow{4}{*}{FFN}
                     & LoRA            & 1106 MB  & 108.5 ms \\
                     & LoRA (compiled) & 1106 MB  & 107.9 ms \\
                     & SPT (3/4)       & 993 MB   & 84.6 ms  \\
                     & SPT (1/2)       & 928 MB   & 54.9 ms  \\
              \bottomrule
       \end{tabular}
       % 336 mm, 506 mm
       \bigskip
       \subcaption{LLaMA-4096}
       \begin{tabular}{cccc}
              \toprule
              Module & Method          & Peak Mem & Duration \\
              \midrule
              \multirow{4}{*}{MHA}
                     & LoRA            & 3799 MB  & 205.2 ms \\
                     & LoRA (compiled) & 3478 MB  & 179.8 ms \\
                     & SPT (1/4)       & 3253 MB  & 238.4 ms \\
                     & SPT (1/8)       & 2711 MB  & 185.2 ms \\
              \midrule
              \multirow{4}{*}{FFN}
                     & LoRA            & 3329 MB  & 300.1 ms \\
                     & LoRA (compiled) & 3502 MB  & 296.6 ms \\
                     & SPT (3/4)       & 2972 MB  & 228.9 ms \\
                     & SPT (1/2)       & 2811 MB  & 150.8 ms \\
              \bottomrule
       \end{tabular}
\end{table}

\begin{table*}[!t]
	\small
	\centering
	\setlength{\tabcolsep}{2.0mm}
	\caption{Breaking the computation time down to kernels for MHA and FFN.}
	\label{tab:eval-breakdown}
	\begin{tabular}{c|c|c|c|c|c}
		\toprule
		\textbf{Method} & \textbf{Part}                                                                              & \textbf{Description}                                       & \textbf{Kernel}                 & \textbf{Duration} & \textbf{Ratio}  \\
		\midrule
		\multirow{7}{*}{\textbf{LoRA}}
		       & \multirow{4}{*}{MHA}
		       & LoRA matrix decomposition and multi-head feature transformation
		       & sgemm\_128x64\_tn                                                                 & 13.5 ms                                           & 64.6\%                                     \\
		\cmidrule(lr){3-6}
		       &                                                                                   & \multirow{2}{*}{Multi-head attention calculation}
		       & sgemm\_128x128\_nn                                                                & 2.8 ms                                            & 13.1\%                                     \\
		       &                                                                                   &                                                   & softmax\_warp\_forward & 0.9 ms   & 4.2\%  \\
		\cmidrule(lr){3-6}
		       &                                                                                   & Reshape, addition, or activation operators
		       & elementwise\_kernel                                                               & 1.3 ms                                            & 6.0\%                                      \\
		\cmidrule(lr){2-6}
		       & \multirow{3}{*}{FFN}
		       & \multirow{2}{*}{LoRA matrix decomposition and feedforward feature transformation}
		       & sgemm\_128x64\_tn                                                                 & 13.8 ms                                           & 47.6\%                                     \\
		       &                                                                                   &                                                   & sgemm\_128x128\_tn     & 13.8 ms  & 47.6\% \\
		\cmidrule(lr){3-6}
		       &                                                                                   & Reshape, addition, or activation operators
		       & elementwise\_kernel                                                               & 1.1 ms                                            & 3.9\%                                      \\
		\midrule
		\multirow{9}{*}{\textbf{SPT}}
		       & \multirow{4}{*}{MHA}
		       & LoRA matrix decomposition and multi-head feature transformation
		       & sgemm\_128x64\_tn                                                                 & 13.5 ms                                           & 61.9\%                                     \\
		\cmidrule(lr){3-6}
		       &                                                                                   & \multirow{3}{*}{Multi-head attention calculation}
		       & cusparse::sddmm\_ker                                                              & 1.8 ms                                            & 8.4\%                                      \\
		       &                                                                                   &                                                   & cusparse::csrmm\_alg2  & 1.7 ms   & 8.0\%  \\
		       &                                                                                   &                                                   & pq\_lookup\_kernel     & 1.1 ms   & 5.0\%  \\
		\cmidrule(lr){2-6}
		       & \multirow{5}{*}{FFN}
		       & \multirow{2}{*}{LoRA matrix decomposition and feedforward feature transformation}
		       & sgemm\_128x64\_tn                                                                 & 7.1 ms                                            & 48.4\%                                     \\
		       &                                                                                   &                                                   & sgemm\_128x64\_nn      & 6.9 ms   & 47.1\% \\
		\cmidrule(lr){3-6}
		       &                                                                                   & \multirow{2}{*}{Index select or put operators}
		       & index\_put\_kernel                                                                & 1.0 ms                                            & 6.8\%                                      \\
		       &                                                                                   &                                                   & index\_get\_kernel     & 0.4 ms   & 2.7\%  \\
		\cmidrule(lr){3-6}
		       &                                                                                   & Reshape, addition, or activation operators
		       & elementwise\_kernel                                                               & 0.7 ms                                            & 4.8\%                                      \\
		\bottomrule
	\end{tabular}
\end{table*}

\stitle{Time and memory costs}
In this experiment, we look into the time and memory consumption of MHA and FFN to validate the effectiveness of our optimizations.
We adopt the same batch size and sequence length settings as in \ref{sec:eval-overview} and report the results in Table~\ref{tab:eval-profile}.
Entries marked with \textit{compiled} means that they are complied with JIT, an option provided by the latest PyTorch.
For \name{}, the numbers in the brackets are the sparsity strengths used for MHA and FFN, and we set them as practical values that yield good model quality. We make the following observations from Table~\ref{tab:eval-profile}.

First, the sparse MHA effectively reduces the peak memory consumption compared with LoRA, and the reduction is larger with stronger sparsity. Specifically, considering OPT-2048, the reduction is 32\% with a non-zero portion of 1/4 and 57\% with 1/8 non-zeros.
This, as explained in Section~\ref{sec:algorithm}, sparse MHA only stores the top attention weights while standard MHA stores the attention weights between all token pairs, and stronger sparsity means that fewer attention weights are stored.
Note that the sparse MHA has lower time complexity compared to vanilla MHA as shown in \S\ref{subsec:sparse MHA}, but provides limited speedup on GPUs. This is because modern GPUs are optimized for contiguous memory access patterns, while sparse attention involves more irregular memory accesses, which limit the actual speedup gained.

Second, our routed FFN effectively reduces the running time compared with vanilla FFN. This is accomplished by the routed FFN reducing the number of parameters activated per token. Additionally, Table~\ref{tab:eval-profile} demonstrates that the speedup achieved by the routed FFN is near the theoretical maximum, i.e., achieving 2.0x and 1.3x speedups with non-zero portions of 1/2 and 3/4, indicating our implementations utilize computational resources efficiently.
The peak memory reduction brought by routed FFN is less significant because the sizes of the input, output, and weight tensors remain unchanged.
Moreover, compiled LoRA yields limited gain over LoRA in both time and memory. This is because the baseline has already been highly optimized, with most of the time spent on basic matrix calculations that are difficult to optimize further.
This shows that it is necessary to conduct tailored algorithm and system optimizations for Transformer as in our work.

\stitle{Kernel breakdown}
In Table~\ref{tab:eval-breakdown}, we break the running time of our sparse MHA and routed FFN into their constituting GPU kernels and compare them with the vanilla counterparts.
The Transformer block is OPT-2048, and we consider only the forward pass for simplicity. We make the following observations from the results.

First, peak memory reduction is the priority for sparse MHA since the LoRA decomposition and multi-head feature transformations inevitably take up most of the compute time, which leaves limited opportunity for speedup.
The high overhead of multi-head feature transformation was also observed in other research~\cite{shazeer2019multiquery}.
Moreover, sparse MHA offers no speedup over dense matrices because dense matrix operators enjoy fast contiguous memory accesses from modern hardware while random memory accesses occupy a large part of sparse matrix operators.
According to these observations, we decide that reducing the memory usage of MHA should take priority over achieving speedup.
Second, routed FFN effectively reduces the computation cost of FFN (i.e., which are mainly GEMM operations), and the overhead of the router network is negligible.
The dynamically skipped blocks in the routed FFN translate almost directly to an equivalent speedup.
The breakdown also shows that our routed FFN introduces \textsf{index\_put} and \textsf{index\_get} operators for batching tokens, and their overheads are only 13\% of overall running time.

\begin{table}[!t]
    \small
    \centering
    \setlength{\tabcolsep}{3.5mm}
    \caption{Compare with an alternative MHA implementation.}
    \begin{tabular}{cccc}
        \toprule
        \textbf{Module} & \textbf{Method} & \textbf{Peak Mem} & \textbf{Duration} \\
        \midrule
        \multirow{2}{*}{MHA}
                        & SPT             & 1123 MB           & 54.1 ms           \\
                        & Naive-PQ        & 1253 MB           & 248.9 ms          \\
        \bottomrule
    \end{tabular}
    \label{tab:eval-alternative}
\end{table}

\stitle{Alternative implementations}
In Table~\ref{tab:eval-alternative}, we compare our sparse MHA with an alternative implementation, which also uses PQ but adopts the standard practice to compute and sort floating point query-key similarity scores. The Transformer block is OPT-2048, and the experiment settings are the same as Table~\ref{tab:eval-profile}. Table~\ref{tab:eval-alternative} shows that the Naive-PQ implementation has slightly higher memory consumption than our bucket-sort based implementation but its running time is $4.6\times$ of our implementation. This is because Naive-PQ suffers from a high complexity in computing and sorting floating point query-key similarity scores, and shows the effectiveness of our implementation.

We also experimented with using the BSR approach as an alternative for our routed FFN, which generates a parameter mask for each token to conduct batched computation. This experiment runs OOM because when the input tokens are the size of [16, 512], the BSR masks take up 200GB, which far exceeds our GPU memory capacity. In contrast, our routed FFN implementation batches the tokens according to their activated parameter blocks. It does not rely on the masks and achieves a speedup close to theory as shown by Table~\ref{tab:eval-alternative}.
Implementing the proposed optimizations is critical to enabling \name{} and avoiding system failures due to excessive memory and computation demands.

% Related Work
\section{Related Work}
\label{sec:related}

Large Transformer models, such as BERT~\cite{devlin2018bert}, RoBERTa~\cite{liu2019roberta}, GPT-2~\cite{radford2019gpt2}, OPT~\cite{zhang2022opt}, LLaMa~\cite{touvron2023llama}, and ViT~\cite{dosovitskiy2021vit}, achieve tremendous success recently and are used for an increasing number of applications.
However, these models are prohibitively expensive to train due to their large sizes, and we discuss the most widely used techniques for training or fine-tuning large Transformer models.

\stitle{Distributed training}
To handle large datasets and models, Transformer training can be parallelized across many GPUs. As the two most common approaches, data parallelism assigns the GPUs to process different training samples by keeping a replica of the model on each GPU, and model parallelism partitions the model over the GPUs and transfers intermediate activation across GPUs~\cite{li20torchddp}.
Pipeline parallelism reduces the idle time of model parallelism by pipelining multiple mini-batches~\cite{huang2019gpipe}, and hybrid parallelism flexibly partitions training workload in dimensions such as sample, feature, and model for short training time~\cite{kim19parallax}. Many systems support distributed training. For instance, DeepSpeed~\cite{rasley20deepspeed} and Megatron-LM~\cite{narayanan21megatron} optimize pipelining and scheduling across multiple GPUs for training throughput. PyTorch introduces fully sharded data parallelism (FSDP)~\cite{zhao2023fsdp} to simplify distributed training for large Transformers. Our work is largely orthogonal to these distributed training works, and the only consideration is that when the distribution strategy partitions computations intra-layer, the mentioned operators in \S\ref{sec:system} would need distributed implementations.

\stitle{Accelerate Transformers}
Some works conduct fine-grained CPU or GPU optimizations for Transformer execution. FasterTransformer and FlashAttention~\cite{dao2022flash} implement hand-crafted GPU kernels to minimize data transfer and reuse data in cache.
TurboTransformers~\cite{fang2021turbo} shows that reduced memory footprint and better memory management can reduce the running time of Transformer models.
Triton~\cite{tillet19triton} is utilized by PyTorch as a compiler to optimize the computation graphs of neural networks, it achieves comparable performance to hand-tuned kernels, which served as part of our baselines for evaluating time and memory costs in \S\ref{sec:eval-ablation}.
Model quantization converts the floating point model weights to lower precision to improve efficiency but model quality is usually sacrificed~\cite{dettmers22int8, dettmers23qlora}. Our GPU kernel optimizations target the sparse MHA and routed FFN and thus are different from these works, and we could also benefit from using low-precision model weights.

\stitle{Fine-tuning Transformers}
Adapter-based methods are widely used to fine-tune Transformers for downstream tasks~\cite{houlsby19transfer, chen2022adapter}. They freeze the parameters of the pre-trained model to preserve the learned knowledge and introduce a small number of trainable parameters to adapt to the specific task.
In particular, BitFit~\cite{zaken22bitfit} only fine-tunes the bias vectors in a pre-trained model. Input-Tuning~\cite{an22inputadapter} introduces adapters in the input embedding layers to process contexts. LoRA~\cite{dettmers23qlora} adds low-rank matrices to all layers and yields competitive model quality to full tuning.
Transformer is also widely used by computer vision models recently~\cite{dosovitskiy2021vit}.
A popular approach to fine-tune vision Transformers is called ControlNet~\cite{zhang2023controlnet}, which adds convolution control networks to a pre-trained model. Our work improves LoRA, the state-of-the-art fine-tuning method for Transformer-based language models, and may also benefit Transformers in computer vision models.

\stitle{Transformer variants}
It has been observed that the output of the softmax function is dominated by a few top entries for classification models~\cite{martins16sparsemax, zhang18dynclass, zhao21annsoftmax}. Our sparse MHA is driven by similar observation but tackles the challenge of efficiently identifying the top-$L$ query-key pairs.
To process longer sequences, ReFormer~\cite{kitaev2020reformer} replaces the inner product attention with locality sensitive hashing (LSH) attention, and Recurrent Memory Transformer~\cite{bulatov2022recurrent} uses an additional external memory module to store long-term context. Mixture-of-Experts (MoE)~\cite{shazeer2017moe} and Switch~Transformer~\cite{fedus2022switch} introduce sparsely activated weight for Transformers by modifying the FFN module to improve model capacity.
Some of these works share the idea of utilizing sparsity with our work but their purposes are different, e.g., process longer sequence and increase model capacity. Moreover, they are required to train the models from scratch using their Transformer architectures while we focus on fine-tuning pre-trained models.

% Conclusion
\section{Conclusions}\label{sec:conclusion}

In this paper, we present the \name{} system for the efficient fine-tuning of Transformer-based large language models. We observe that the main memory consumption of Transformer is used to store the attention weights for multi-head attention (MHA) and the majority of running time is spent on the feed-forward network (FFN). From the algorithm perspective, we propose the sparse MHA and routed FFN modules to expose sparsity for reduced memory consumption and running time. From the system perspective, we customize CUDA implementations for sparse MHA and routed FFN to conduct sparse training efficiently on GPU. Experiment results show that \name{} effectively both reduces training time and memory consumption.

%\begin{acks}
%  To Robert, for the bagels and explaining CMYK and color spaces.
%\end{acks}

%\clearpage
\bibliographystyle{ACM-Reference-Format}
\bibliography{sample-base}

% Appendix
\clearpage
\appendix

\section{Artifact and Reproducibility}

This appendix provides detailed guidance on replicating the most significant experimental outcomes obtained by \name{}.
In \S\ref{sec:appendix-setup}, we present the experimental setup. We then show three main topics: correctness~(\S\ref{sec:appendix-correct}), efficiency~(\S\ref{sec:appendix-efficiency}), and model quality~(\S\ref{sec:appendix-quality}).

\subsection{Experiment Setup}\label{sec:appendix-setup}

Preparation procedures for the experiments.

\textbf{Hardware.}
This experiment is configured to be runnable on a single NVIDIA RTX 3090 (24GB memory) GPU.
We also tested RTX 2080 TI (11 GB memory). Other GPU models may work, but have not been fully validated.

\textbf{Software.}
\begin{table}[h]
    \small
    \centering
    \setlength{\tabcolsep}{5.0mm}
    \caption{Software versions.}
    \begin{tabular}{cc}
        \toprule
        name         & version             \\
        \midrule
        Ubuntu       & 18.04 or above      \\
        Kernel       & 4.15.0-175 or above \\
        GCC          & 7.5.0 or above      \\
        GPU-Driver   & 465.19.01 or above  \\
        CUDA-Toolkit & 11.8, minimum 11.0  \\
        Python       & 3.10.7 or above     \\
        PyTorch      & 2.0.1 or above      \\
        \bottomrule
    \end{tabular}
    \label{tab:appendix-software}
\end{table}
The required softwares are listed in Table~\ref{tab:appendix-software}.

\textbf{Build Essentials.}
\textit{build-essential}, \textit{python3-setuptools}, and \textit{ninja-build} are required.

\textbf{Installing CUDA.}
Download and install CUDA 11.8 and compatible drivers~\footnote{https://developer.nvidia.com/cuda-downloads}.

\textbf{Installing Conda.}
Install Anaconda or Miniconda~\footnote{https://docs.conda.io/projects/miniconda/en/latest/}.

\textbf{Creating Python Environment.}
Create a new clean environment with \textit{``conda create -n spt python=3.10 \&\& source activate spt''}

\textbf{Installing dependencies.}
The following packages are required:
\begin{itemize}
    \item PyTorch Basic: \textit{pip3 install torch==2.0.1 torchdata transformers}
    \item End-to-end: \textit{pip3 install lightning==2.0.8 deepspeed torchmetrics}
\end{itemize}
Note that here we use \textit{pip3} to install packages inside the virtual environment, \textit{conda} can also be used here, but some package names may differ between pip and conda.

\subsection{Correctness}\label{sec:appendix-correct}

Correctness is the most fundamental prerequisite for the whole implementation. We ensure \name{} works as expected by performing extensive unit tests. Here we list 7 main unit tests,

\begin{itemize}
    \item \textit{test\_cdist.py}: checking CUDA kernel results on fused collection distance calculation kernel (\textsf{cdist} in~\S\ref{subsec:sys-mha}) with unfused PyTorch implementation.
    \item \textit{test\_lookup.py}: checking CUDA kernel results of the bucket sort shown in~\S\ref{subsec:sys-mha}, this ensures that \name{} retrieved the query-key vectors correctly.
    \item \textit{test\_sddmm.py} and \textit{test\_spmm.py}: checking sparse matrix multiplications with equivalent dense matrix computations, this also ensures the \textit{Indptr} and \textit{Indices} shown in~\S\ref{subsec:sys-mha} are well constructed.
    \item \textit{test\_softmax.py}: checking CUDA kernel results on sparse matrices. Note that softmax is numerically sensitive, a tiny $error \le 0.01$ is allowed in the unit test.
    \item \textit{test\_sparse\_mha.py} and \textit{test\_routed\_ffn.py}: checking the correctness of the proposed sparse MHA and routed MHA by comparing results with naive and not optimized implementations.
\end{itemize}

\begin{figure}[!t]
    \begin{scriptsize}
        \begin{verbatim}
    def test_softmax():
        # y_1, grad_1: torch kernel
        y_1 = torch.softmax(dense, dim=-1)
        torch.max(y_1).backward()
        grad_1 = dense.grad.detach().clone()
        
        # y_2, grad_2: custom kernel
        y_2 = kernels.softmax(
            indptr, indices, values
        )
        torch.max(y_2).backward()
        grad_2 = values.grad.detach().clone()

        # check results
        assert torch.allclose(y_1, y_2.to_dense(), atol=1e-2)
        assert torch.allclose(grad_1, grad_2.to_dense(), atol=1e-2)
    \end{verbatim}
    \end{scriptsize}
    \caption{A sample unittest to check both forward and backward passes of an operator.}
    \label{fig:appendix-check}
\end{figure}

Note that both forward pass and backward pass results are carefully verified, Figure~\ref{fig:appendix-check} shows a sample unittest test to make sure the customized operators computes correct output and gradient.

\subsection{Efficiency}\label{sec:appendix-efficiency}

We provide a python script \textit{script/profile.py} to easily profile the main modules of \name{}, different Transformer configurations and fine-tuning methods can be set. The script takes 8 arguments:
\begin{itemize}
    \item \textit{--name} (required): specifying the Transformer configuration to use, can be one of \textit{opt-1024}, {opt-2048}, \textit{opt-2560}, {llama-2560}, or {llama-4096}.
    \item \textit{--tuning} (required): choosing the fine-tuning method to evaluate, which can be \textit{full}, \textit{lora}, or \textit{sparse}.
    \item \textit{--module} (required): specifying the module to evaluate, can be one of \textit{mha}, \textit{ffn}, or \textit{both}.
    \item \textit{--d\_lora} (default 16): specifying the low-rank decomposition dimension of LoRA, which is ignored for full-parameter tuning.
    \item \textit{--seq\_length} (default 512) and \textit{--batch\_size} (default 16): specifying the sequence length and batch size.
    \item \textit{--backward}: choosing whether or not to profile on backward passes.
    \item \textit{--compile}: choosing whether or not to enable \textit{torch.compile}.
\end{itemize}

\begin{figure}[!t]
    \begin{scriptsize}
        \begin{verbatim}
        # logs by our Model Adapter
        [UPGRADE] mha.linear_q Linear -> LoRALinear
        [UPGRADE] mha.linear_k Linear -> LoRALinear
        [UPGRADE] mha.linear_v Linear -> LoRALinear
        [UPGRADE] mha.linear_o Linear -> LoRALinear
        [UPGRADE] ffd.fc1 Linear -> LoRALinear
        [UPGRADE] ffd.fc2 Linear -> LoRALinear

        # breakdown and timing (20 runs)
        -------------------------------------------------------------------
        Name      Self CUDA   Self CUDA %    CUDA total 
        -------------------------------------------------------------------
        ampere_sgemm_64x64_tn            1.216s        37.55%     1.216s
        ampere_sgemm_128x32_nn           624.253ms     19.27%     624.253ms
        ampere_sgemm_128x64_tn           290.456ms      8.97%     290.456ms
        ampere_sgemm_64x64_nn            267.770ms      8.27%     267.770ms
        Memcpy DtoD (Device -> Device)   109.602ms      3.38%     109.602ms
        -------------------------------------------------------------------
        Self CPU time total: 3.004s

        # peak memory statistics
        |======================================================================|
        |                PyTorch CUDA memory summary, device ID 0              |
        |----------------------------------------------------------------------|
        |           CUDA OOMs: 0          |        cudaMalloc retries: 0       |
        |======================================================================|
        |     Metric       | Cur Usage  | Peak Usage | Tot Alloc  | Tot Freed  |
        |----------------------------------------------------------------------|
        | Allocated memory | 416936 KiB |   2714 MiB | 449859 MiB | 449452 MiB |
        | from large pool  | 409856 KiB |   2704 MiB | 449168 MiB | 448768 MiB |
        | from small pool  |   7080 KiB |     11 MiB |    691 MiB |    684 MiB |
        |----------------------------------------------------------------------|
    \end{verbatim}
    \end{scriptsize}
    \caption{A sample output of the profile script.}
    \label{fig:appendix-output}
\end{figure}

Specifically, \name{} can be reproduced by running: \textit{python3 script/profile.py --name='opt-2048' --tuning='sparse' --module='both'}.

The output is printed into the \textit{stdout} stream, which includes all the important metrics and statistics.
Figure~\ref{fig:appendix-output} shows a sample profiling output. Specifically, system configurations, kernel time consumption, and memory statistics are printed in sequence. Note that memory statistics are generated on top of PyTorch's buffer pool, thus there may be subtle differences under different GPU driver and PyTorch version.

\subsection{Model Quality}\label{sec:appendix-quality}

In this section, we present approaches for assessing the model quality fine-tuned by \name{} from two perspectives. Firstly, we evaluate the MMLU score by loading our fine-tuned parameters. Secondly, we provide an end-to-end script that replicates the training procedures.

\begin{table}[h]
    \small
    \centering
    \setlength{\tabcolsep}{2.0mm}
    \caption{Required checkpoints.}
    \begin{tabular}{cccc}
        \toprule
        model & file                        & size     & checksum \\
        \midrule
        \multirow{2}{*}{OPT-125M}
              & opt-125m.ckpt               & 642 MB   & 274e1ea1 \\
              & opt-125m-spt.ckpt           & 17 MB    & 77a96766 \\
        \multirow{2}{*}{OPT-2.7B}
              & opt-2.7b.ckpt               & 10622 MB & 05d331f6 \\
              & opt-2.7b-spt.ckpt           & 100 MB   & d6d5c56b \\
        \multirow{2}{*}{LLaMA-2.7B}
              & sheared-llama-2.7b.ckpt     & 10434 MB & 8cac7a74 \\
              & sheared-llama-2.7b-spt.ckpt & 102 MB   & 67f00f2c \\
        \bottomrule
    \end{tabular}
    \label{tab:appendix-checkpoints}
\end{table}

\textbf{MMLU evaluation.}
We offer three checkpoints (shown in Table~\ref{tab:appendix-checkpoints}, checksum is the leading 8 hexdigest of \textit{sha256sum}) for the fine-tuned model, which validate the quality of the reported results.
The MMLU evaluation pipeline operates as follows:
\begin{itemize}
    \item Loading pre-trained model weights from the HuggingFace~\footnote{https://huggingface.co/facebook/opt-125m, https://huggingface.co/facebook/opt-2.7b, https://huggingface.co/princeton-nlp/Sheared-LLaMA-2.7B} model hub and converting them to PyTorch format: \textit{python3 script/profile.py --name='facebook/opt-2.7b'}.
    \item Loading our fine-tuned weights (the smaller files show in Table~\ref{tab:appendix-checkpoints}) that patch onto the pre-trained weights, and evaluating the MMLU score: \textit{python3 script/mmlu-evaluate.py --ckpt='opt-2.7b.ckpt' --spt\_ckpt='opt-2.7b-spt.ckpt'}.
\end{itemize}
This script only evaluates MMLU on 64 samples by default, which can be set to higher via \textit{--test\_batches=1024}.

\textbf{End-to-end fine-tuning.}
Additionally, we provide end-to-end fine-tuning scripts to replicate the model fine-tuning process.
The \textit{sparse-tuning-0.py} script is configured to utilize a single GPU and train the smallest 125M model. On the other hand, \textit{sparse-tuning-1.py} leverages deepspeed for distributed capabilities and requires 4 GPUs.
The novel modules mentioned in this work are all plugable. By customizing the given scripts, models can be fine-tuned while retaining only sparse MHA or routed FFN module.

\end{document}